\documentclass[twocolumn,dvipsnames]{aastex63}
\usepackage[T1]{fontenc}
\usepackage{amsmath}
\usepackage{natbib}
\usepackage{soul}
\usepackage{xcolor}
\usepackage{lipsum}
\RequirePackage{lineno}
\usepackage{graphicx}	
\usepackage{multirow,makecell}

\RequirePackage{bm,amsfonts,hhline,amsmath,amssymb,microtype,float,eurosym,
latexsym,epsf,mathtools,cuted,times,makecell,array,natbib}
\usepackage{tabularx}
\usepackage{epstopdf}

\usepackage{gensymb}
\usepackage{xcolor}

\usepackage{pifont}

\usepackage{hyperref}

\hypersetup{colorlinks=true, linkcolor=red, filecolor=magenta, urlcolor=Sepia, citecolor=blue}

\definecolor{ForestGreen}{HTML}{228B22}

\received{XXX}
\revised{YYY}
\accepted{ZZZ}

\submitjournal{ApJ}

\shorttitle{Continuous gravitational waves from neutron stars}
\shortauthors{Das \& Mukhopadhyay}

\begin{document}

\title{Detection possibility of continuous 
gravitational waves from rotating magnetized neutron stars}

\author{Mayusree Das$^{\dagger}$}
\email{mayusreedas@iisc.ac.in$^{\dagger}$}
\affiliation{Joint Astronomy Programme, Department of Physics, Indian Institute of Science, Bangalore 560012, India}

\author[0000-0002-3020-9513]{Banibrata Mukhopadhyay$^{\ddagger}$}
\email{bm@iisc.ac.in$^{\ddagger}$}
\affiliation{Joint Astronomy Programme, Department of Physics, Indian Institute of Science, Bangalore 560012, India}

\begin{abstract}
In the past decades, several neutron stars (NSs), particularly pulsars, with mass $M>2M_\odot$, have been 
	observed. On the other hand, the existence of massive white dwarfs (WDs), even violating the Chandrasekhar
	mass limit, was inferred from the peak luminosities of type Ia supernovae. Hence, there is a generic
	question of the origin of massive compact objects. Here we explore the existence of massive, 
	magnetized, rotating NSs with the soft and steep equation of states (EoSs) by solving axisymmetric 
	stationary stellar equilibria in general relativity. For our purpose, we consider the Einstein
	equation solver for stellar structure {\it XNS} code. Such rotating NSs with magnetic field and 
	rotation axes misaligned, hence with non-zero obliquity angle, can emit continuous gravitational waves (GW), which can be detected by 
	upcoming detectors, e.g., Einstein Telescope, etc. We discuss the decay of the magnetic field, angular
	velocity, and obliquity angle with time due to angular momentum extraction by GW and dipole radiation,
 which determine the timescales related to the GW emission. 
	{Further, in the Alfv\'en timescale, a differentially rotating, massive proto-NS rapidly settles 
	into a uniformly rotating, less massive NS due to magnetic braking and viscosity.}
	These explorations suggest that detecting massive NSs is challenging and sets a timescale for 
	detection. We calculate the signal-to-noise ratio of GW emission, which confirms that any detector
	cannot detect them immediately, but detectable by Einstein Telescope and Cosmic Explorer over months of integration time, 
	leading to direct detection of NSs.
\end{abstract}


\section{Introduction}

Recently gravitational wave (GW) has been detected from several binary mergers by LIGO and VIRGO, 
when either both the components are black hole (BH) or both are neutron star (NS). Also, there is an event
GW190814, when one is a BH, and the other could be either a massive NS or a lighter BH. All these events 
have enlightened GW astronomy \citep{AB2016B,AB2017A,AB2017B,AB2017C,AB2017D,LV2018}. 
However, isolated objects may also 
emit continuous GW (CGW) at a certain frequency \citep{ZS1979}, 
if, e.g., they have mountains and holes around the surface, misaligned magnetic and rotation axes (obliquity angle non-zero).
For different possible mechanisms generating CGW, see the review by 
\citet{RR2017}. In this paper, we consider triaxial sources with a violation of axisymmetry because of non-zero obliquity
angle \citep{BG1996,JA2002,F2010,FS2017,BM2017}.

\citet{KM2019} argued that strong gravitational radiation could be emitted from rotating magnetized NSs and 
white dwarfs (WDs), even with a small obliquity angle, which can be detected by upcoming GW detectors. 
While it is known that compact objects in a binary system radiate more powerful GW than CGW, which, however 
can be detected by at a different frequency range \citep{KM2019}, and hence they can be distinguished. 

NSs are assumed/inferred to be born with mass $M\sim 1.4M_\odot$, on average \citep{Zhang,Miller}, from the evolution of stars with masses $10M_\odot\lesssim M\lesssim20M_\odot$. However, NSs as accreting millisecond pulsars have higher masses, $M\sim 1.6M_\odot$, on average \citep{Zhang,MB2022}. Recently, higher masses of binary NS systems like millisecond pulsars PSR J1614-2230: $M=1.97\pm0.04M_\odot$ \citep{DPR2010}, PSR J0348+0432: $M=2.01\pm0.04M_\odot$ \citep{ANT2013}, and PSR J0740+6620: 
$M=2.04_{-0.10}^{+0.09}M_\odot$ \citep{CROM2020} have been measured. Very recently, a new GW event reported by Ligo-Virgo collaboration as GW190814 \citep{ABBOTT2020} was observed with a $22.2-24.3M_\odot$ black hole and a $2.50-2.67M_\odot$ secondary component. As the mass of the secondary component of GW190814 lies in the lower mass region of the mass gap ($2.5M_\odot<M<5M_\odot$), it can be a candidate for very massive NS. Such high mass NSs can be inferred to be formed due to high magnetic field and rotation \citep{PBDZ2014}. \citet{CHU2014}, \citet{DB2021}, and \citet{R2021} showed that increasing anisotropy and magnetic field strength also can support the existence of massive NSs. It is important to mention that the variance of EoS can generate massive NSs as well \citep{LAT2012}. However, in order to detect the NSs via GW, we need high ellipticity (hence deformation), and thus we necessarily need to introduce NS with a high magnetic field and rotation. Therefore, NSs producing strong CGW are the candidates for 
massive NSs.

Isolated NSs are very lowly luminous 
objects due to their tiny size and no source of energy. Thus, we have to rely upon their other activities
to detect them directly, and one of them is their CGW. Interestingly, there has been no detection of CGW 
from NSs in LIGO, VIRGO, aLIGO, and aVIRGO so far \citep{AB2019J,AB2019JJ,P2020}. If any of them is detected 
in the future, e.g., by Einstein Telescope, Cosmic Explorer, etc., depending on its distance from the Earth, 
their magnetic field can be predicted from ellipticity (the parameter measuring the degree of deformation of 
the object) related to their GW amplitude. Hence, the detection of CGWs from isolated NSs would be a fundamental 
breakthrough, which can provide us with an idea about their spin, magnetic field, mass, as well as EoS.

One of the potential origins of high magnetic fields in a NS, e.g., magnetar, is the frozen flux, particularly if the field is not disrupted during the core-collapse of a massive star \citep{WOLT1964,FMZ2015}. However, a strong fossil field brakes stellar rotation \citep{SCH2017}, hence a fossil field is probably not compatible with the existing fast rotation of NSs. An additional mechanism, like a dynamo \citep{FP1975,MOF1978, TD1993,Brandenburg2001,BRU2003,BRAN2005} or the magnetorotational instability \citep{FRI1969,BH1991,Ober2009,GE2015,MRRC2015,RGO2016}, is needed to enhance this fossil field above $10^{15}$ G. The core of the proto-NS experiences neutrino-driven turbulent convection during the first few tens of seconds \citep{MIRA2000,Thom2001,Miralles_2002}. A strong differential rotation is also present if the proto-NS is generated from the merger or remnant of supernovae. Differential rotation along with convection can generate $\alpha-\Omega$ dynamo acting over the whole NS core \citep{DT1992}. By differential rotation, the frozen poloidal magnetic field is twisted into the toroidal field. However, the total kinetic energy available from the core differential rotation corresponds to a maximum field of $\simeq 10^{17-18}$ G, depending on the initial spin period \citep{TD1993,Dun1998}. In slowly rotating NSs, differential rotation is, however, negligible, and a convection-driven $\alpha$-type dynamo results in a much weaker large-scale field \citep{DT1992,TD1993}.

The rotational frequency and obliquity angle of rotating magnetized NSs decay with time due to the extraction 
of angular momentum by GW and dipole radiation. 
The spin-down timescale or age was calculated for spherically symmetric stars considering the decays of spin and the obliquity angle simultaneously of the 
pulsar having dipole radiation long back by \citet{MG1970} and \citet{DG1970}.
Then the quadrupolar radiation was included in the calculation along with the 
dipole radiation \citep{CH1970}. The equations were generalized by \citet{MELATOS2000} for non-axisymmetric 
stars because magnetic field and rotation actually deform the star. Also, this formalism has been used to 
describe the highly magnetized NSs known as magnetars \citep{LZLL2018,SCA2019,LJ2020}. 

{Nevertheless, the viscous effect of NS matter may try to
increase the obliquity angle by redistributing angular momentum. This may turn the NS into an orthogonal rotator, if the NS is a toroidal magnetic field dominated. The early evolution of the obliquity angle depends on the relative strength between dipole/quadrupole radiations and viscous damping. If the viscous damping is dominating for a particular star/model, the star will be able to radiate GW for a longer time \citep{CC2002,Dall2009}.}

However, the magnetic field also decays 
due to Ohmic dissipation, ambipolar diffusion, and Hall drift \citep{GR1992,HK1998}, shown for magnetar. Each of 
these processes may dominate the decay depending on the magnetic field strength. \citet{MON2021} studied 
spin-down and magnetic field decay for NSs to measure the age of present-day magnetars. 

It is important to note that pulsating 
highly magnetized NSs may not radiate GW for very long. Due to the decay mentioned above, after some time, 
either the magnetic and rotation axes align with each other, or they
stop rotating, thus do not behave as pulsars anymore. 
Also, the magnetic field decays, so the ellipticity does, hence the GW radiation also decreases. 
Therefore, it is necessary to study the timescale for which the NS will be detectable by GW. 
This exploration, not performed earlier rigorously to the best of our 
knowledge, suggests that the detection of NS by its CGW may be challenging, depending on
the decay of various parameters. 

We plan to study the plausible instantaneous GW detection from isolated NSs by some upcoming 
detectors. At present, immense effort is going on to increase the sensitivity of GW detectors for various sources \citep{SB2019}, which can be done by improving the signal-to-noise ratio (SNR) of the detector. 
By calculating SNR for a certain integrated time, we can detect GW from a source that could not be detected
instantaneously. Thereby we can estimate the necessary observation time for the particular GW detector 
to detect these objects before their radiation decays away. 

To radiate electromagnetic dipole radiation, 
the NS should have a certain amount of poloidal magnetic field \citep{SOU2020}.
However, \citet{W2014} argued that the NS should eventually be toroidally dominated due to the action of 
$\Omega$-dynamo. Therefore, in this work, we calculate the SNR of GW signal considering toroidally dominated 
NSs. Nevertheless, we will show some results for the poloidally dominated counterpart for completeness. 
Highly magnetized NSs may also be hypermassive due to 
differential rotation acquired at their birth \citep{SU2000,ZM1997,RMR1998}. However, in the 
Alfv\'en timescale, the differentially rotating NS is expected to rapidly settle into a uniformly rotating, 
less massive NS due to magnetic braking \citep{SHAP2000,CSS2003,LS2003}. Note that dynamo driven by 
differential rotation may also be the source of the strong magnetic field of a NS \citep{W2014}.

The specific plan of the present work is the following. One immediate question is: is there any
shortcoming in the underlying models or any specific observational/detectional limitations or 
combination of them? In this paper, we explore the time varying properties of isolated NSs to 
understand their detection possibility in CGW.
We plan to study the time evolutions of magnetic field, rotational frequency and obliquity angle 
simultaneously of isolated NSs. \citet{KMMB2020} showed that rotational frequency 
and obliquity angle of WDs decay with time and hence WDs eventually stop emitting GW.
In this paper, we show that rotating, magnetized NSs with softer EoS can be massive, sometimes 
with $M>2M_\odot$, some of which will be detectable by upcoming detectors. We argue that the non-detection 
of those NSs will put an upper bound on ellipticity and thus magnetic field. 
Further, we calculate the corresponding SNR to detect such objects by various GW detectors in one year 
of integration time. 

In Section \ref{sec:GWmodel}, we discuss the model of the pulsating, magnetized compact object radiating GW. 
Next, in Section \ref{sec:GWamp}, 
we discuss our results for simulated NSs considering various central densities and magnetic field geometries with the change of angular 
frequency. In Section \ref{sec:Bdecay}, 
we discuss how GW amplitude decreases due to magnetic field decay.
The next part, Section \ref{sec:omchaidecay}, is to discuss the timescales of various pulsating NSs behaving as 
pulsars and shows how GW amplitude decays with time. {For the completeness of discussions of various decay channels, Section \ref{sec:vist} explores the viscous and thermal effects in controlling the pulsations.} In Section \ref{sec:snr}, we calculate corresponding 
SNR for poloidally and toroidally dominated NSs and discuss the timescales for detecting them. 
In Section \ref{sec:alfven} we calculate the 
Alfv\'en timescale in which NSs become less massive due to magnetic braking. Finally we end with conclusions in Section \ref{sec:conclusion}.

\section{Modelling gravitational waves from pulsating compact stars}
\label{sec:GWmodel}
\begin{figure}
	\includegraphics[width=\columnwidth]{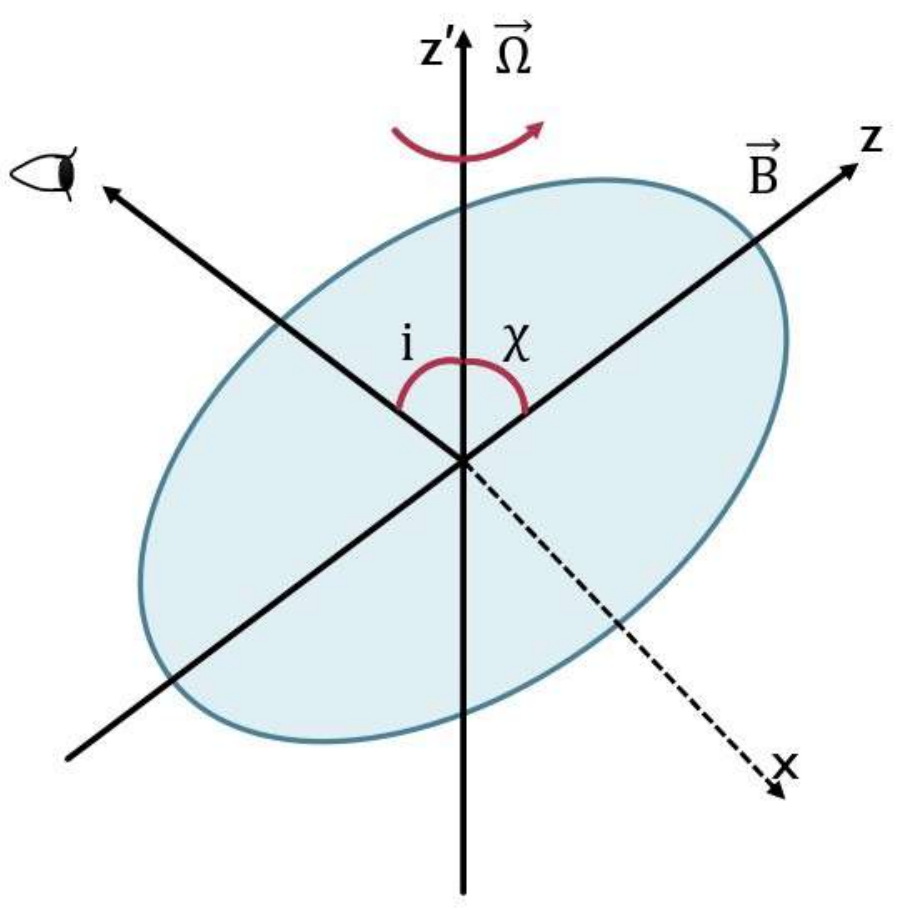}
    \caption{ A cartoon diagram of magnetized rotating compact star with misalignment between the magnetic field
axis and rotation axis.}
    \label{fig:cartoonns}
\end{figure}

It has already been shown that to emit CGW, a compact star must have a non-zero obliquity angle \citep{BG1996}. Fig. \ref{fig:cartoonns} shows a cartoon diagram of a pulsar with $z^\prime$ being the rotation axis and $z$ the magnetic field axis, where the angle between these two axes is $\chi$ (obliquity angle). 
If the magnetic field and rotation are present simultaneously, with misalignment between their respective axes, they comprehensively make the star triaxial, which can produce dipole as well as gravitational radiations. The strain of the two polarizations of the GW at any time $t$ is given by \citep{BG1996,ZS1979}

\begin{eqnarray}
&&h_+ = h_0 \sin \chi \left[ \frac{1}{2} \cos i \sin i \cos \chi \cos \Omega t \right. \\
&&\nonumber \left . - \frac{1+\cos^2 i}{2} \sin \chi \cos 2 \Omega t \right], \\
\nonumber
&&h_\times = h_0 \sin \chi \left[ \frac{1}{2} \sin i \cos \chi \sin \Omega t - \cos i \sin \chi \sin 2 \Omega t \right].\\
\label{eq:gwstrain}
\end{eqnarray}
Here,

\begin{equation}
h_0=\frac{2G}{c^4} \frac{\Omega^2 \epsilon I_{xx}}{d}\left(2\cos^2\chi-\sin^2\chi\right),
\label{eq:gwampn0}
\end{equation}
for  $\chi \to 0$ (but $\neq 0$),

\begin{equation}
h_0=\frac{4G}{c^4} \frac{\Omega^2 \epsilon I_{xx}}{d},
\label{eq:gwamp}
\end{equation}
where $c$ is the speed of light, $G$ is Newton's gravitational constant, $\Omega$ is the angular frequency of the object, $d$ is the distance between the detector and the source object, $i$ is the angle between the rotation axis of the object and our line of sight, and ellipticity is defined as $\epsilon=|I_{zz}-I_{xx}|/I_{xx}$, where $I_{xx}$ and $I_{zz}$ are the principal moments of inertia of the star about $x$- and $z$-axes, respectively. An object behaving as a pulsar can emit CGWs at two frequencies, $\Omega$ and $2\Omega$.

To calculate the quantities such as $I_{xx}$, $I_{zz}$, etc., we use the {\it XNS} code\footnote{https://www.arcetri.inaf.it/science/ahead/XNS/code.html}, a numerical code to 
solve the structure of NSs in general relativity \citep{PBDZ2014}. This code, however, solves only for the 
axisymmetric equilibrium configuration of stellar structure, with rotation (uniform or differential) and/or magnetic field (toroidal or poloidal or mixed field). 
Otherwise, the code solves the time-independent general relativistic magnetostatic equations. 
To assure the stability of a NS, the ratio of magnetic to gravitational energies (ME/GE) to be $\lesssim 10^{-3}$ \citep{KEH1989,BR2009,ARMM2013,HK2017}. However, this limit may be eased up by building a suitable configuration with {a mixed field, which turns out to be toroidally dominated \citep{W2014}.} Nevertheless, our current endeavour is not to study the stability analysis but instead to explore the detectability of GW from NSs with given magnetic field strengths. 
The {\it XNS} code does not have, {as of now}, a provision of a rotating star with appropriate {(i.e., toroidally dominated) and/or an equal fraction of mixed field configuration. It can deal with only poloidally dominated mixed field configuration (twisted torus).} {Therefore,} purely poloidal or purely toroidal magnetic fields, maintaining the ME/GE limit given by \citet{BR2009} mentioned above, should be a valid approximation for poloidally dominated or toroidally dominated mixed field configurations, respectively.

For the solution of a NS, we need to supply the EoS. However, the {\it XNS} code requires
EoS in the polytropic form, i.e., $\mathcal{P}=K\rho^\Gamma$ with  $\mathcal{P}$ being the pressure, 
$\rho$ the density, $\Gamma$ is the polytropic index, and $K$ is the polytropic constant. Nevertheless, the exact 
EoS of NSs is not well established. \citet{PBDZ2014} assumed $\Gamma=2$ and $K=1.45\times10^5$ cm\textsuperscript{5}g\textsuperscript{-1}s\textsuperscript{-2}.
From the concept of tidal deformability, EoS, however, can be constrained from the observation of GW emission \citep{AB2017,AB2018,AB2019} detected by the LIGO/Virgo Collaboration. Based on this, \citet{CH2020} and \citet{DB2021} argued that $\Gamma$ should not be 
$\geq 2$ and EoS to be a bit softer. Therefore, we will choose $\gamma=1.95$, $K=11.3\times10^5$ cm\textsuperscript{5}g\textsuperscript{-1}s\textsuperscript{-2} for this work. Furthermore, if one fits the data 
of actual EoS with the polytropic law, most are well fitted with the polytropic index $\sim 1.8-2.2$. Hence, our choice is justified. {Indeed, it is known that EoS in the polytropic form appropriately captures the ellipticity of NSs as it does for realistic EoS \citep{GCF2011}}. We choose maximum central density $\rho_c=10^{15}$ g cm\textsuperscript{-3} for NS, because above this $\rho_c$, quarks may be produced at the centre due to phase transition, making it a hybrid star.

Moreover, the code implicitly assumes $\chi$ to be zero (or does not include information about $\chi$). However, if there is no misalignment between the magnetic field and rotation axes, the star does not radiate GW radiation. Hence, we make small  $\chi$ approximation in our computations related to CGW based on the {\it XNS} outputs to avoid ambiguity in the structure of the object. 
When $\chi$ is small, the principle moment of inertia of the body about the $z^\prime$-axis, $I_{z^{\prime}z^{\prime}}$, is given by
\begin{equation}
I_{z^{\prime}z^{\prime}}=I_{zz} \cos^2 \chi + I_{xx} \sin^2 \chi.
\end{equation}
Therefore, for the validity of {\it XNS} output as a small $\chi$ approximated solution, we have to assume 
$I_{zz}\cos^2\chi>>I_{xx}\sin^2\chi$ such that, e.g.,  
\begin{equation}
I_{zz} \cos^2 \chi \sim 100 I_{xx} \sin^2 \chi.
\end{equation}
As $I_{zz}\sim I_{xx}$,
\begin{equation}
\frac{\cos^2 \chi}{ \sin^2 \chi}\sim 100,
\end{equation}
which implies $\chi \sim 5.7^{\circ}$.

However, if we could run an efficient code with appropriately chosen $\chi$, we should be able to generate much higher GW strain as it increases with the angle $\chi$. 
This implies that we can use equation (\ref{eq:gwamp}) effectively. Further, the amplitudes of $h_+$ and $h_\times$ in equation (\ref{eq:gwstrain}) will be suppressed by the other factors present therein. 
For instance, at $\chi=3^\circ$,

\begin{eqnarray}
	\nonumber
	&&\text{max} \left( \sin \chi \left[ \frac{1}{2} \cos i \sin i \cos \chi \cos \Omega t \right. \right.\\ 
	&&\left.\left .- \frac{1+\cos^2 i}{2} \sin \chi \cos 2 \Omega t \right] \right)
= 0.0110297
\label{eq:gwampchi}
\end{eqnarray}
for $t=0$ and $i=i_{max}\approx46.5^\circ$. Hence, the maximum amplitude received by the detector at $\chi=3^\circ$ is $h = 0.0110297h_0$, which we consider for further calculations. Further, we assume the distance between the NS and the detector to be 10 kpc. Importantly, only the ellipticity arising due to magnetic field, but not from rotation, is responsible for CGW; hence one has to switch off the rotational effect to extract ellipticity from the run of {\it XNS} code.

\section{Continuous gravitational wave amplitude from massive neutron stars}
\label{sec:GWamp}
We consider purely toroidal and purely poloidal magnetic field cases separately for different EoSs, i.e., with various $\Gamma$, based on the capabilities of \textit{XNS} exploration. In reality, NSs are expected to be consisting of mixed-field geometry. Thus, the actual results might be in between that of purely poloidal and purely toroidal configurations. \citet{KM2019} initiated the variation of $h_0$ for NSs with the change of $\rho_c$, $\Omega$ and magnetic field. Here we explore $h_0$ in order to possibly detect massive NSs and the role of EoS parameters in them.

\subsection{Neutron stars with purely toroidal magnetic field}
\label{sec:Btor}

\begin{figure}
\begin{center}
\includegraphics[width=\columnwidth]{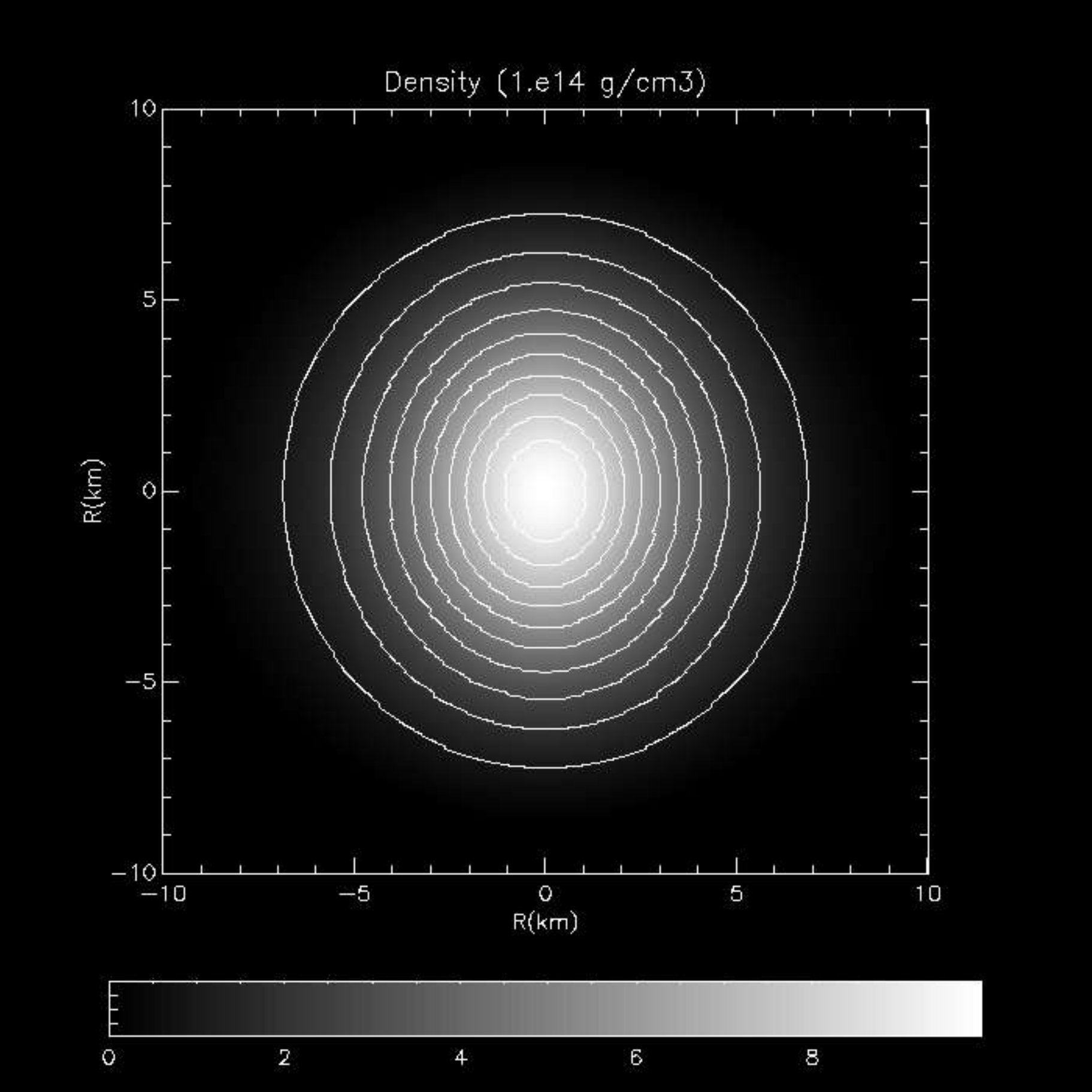}
\caption{Density isocontours of uniformly rotating toroidally magnetized NS of mass $M=1.97M_\odot$ with $\nu=500$ Hz, $B_{max}=4.5\times 10^{17}$ G, ME/GE$=5.6\times 10^{-2}$ and KE/GE$=2.3\times 10^{-2}$.}
\label{fig:torrot}
\end{center}
\end{figure}

It was already known that a purely toroidal magnetic field deforms the NS into a prolate shape and increases its size \citep{PBDZ2014,KM2019}, as seen in Fig. \ref{fig:torrot}. It is observed that the deformation due to the magnetic field at the core is more prominent than in the outer region. However, the rotation makes it oblate; consequently, there is competition between these two opposing 
effects to decide the overall shape of the star. 
In outer layers, where 
the magnetic field decreases and the centrifugal term ($\propto r$ for uniform rotation) becomes large, the shape becomes oblate. For a smaller magnetic field with ME/GE $\lesssim 10^{-3}$, there is practically no deviation of shape from spherical symmetry.

Table \ref{tab:Btor} shows $h_0$ for various uniformly rotating NSs for two $\rho_c$-s 
and a fixed $\Gamma=1.95$. 
Further, for a given $\rho_c$, we consider different magnetic fields $B$ and linear frequency 
$\nu=\Omega/2\pi$.
In the table, $R_E$ is the equatorial radius of NS, $R_P$ the polar radius.
For larger $B$, the NS deviates more from spherical geometry, which produces higher quadrupole moment and, thus, higher ellipticity. Since $h_0\propto\epsilon$, this leads to increasing $h_0$. 
$M$ also increases with $B$ because the star will be able to hold more mass due to increased outward magnetic pressure (but see \citealt{DB2021}). 
We consider different frequencies with higher ME/GE, as well as lower ME/GE ($\leq 10^{-3}$). It shows that NSs with higher ME/GE emit GW more efficiently. From Tables \ref{tab:Btor} and \ref{tab:Bpol} (will be discussed below), it is also clear that highly magnetized NSs are indeed massive, sometimes with $M>2M_\odot$. They should be the targeted to detect by CGW.

\begin{table*}
\begin{center}
\caption{Uniformly rotating NS with toroidal magnetic field for $\chi=3^\circ$. Here $B_{max}$ is the maximum magnetic field when the surface field could be much smaller.}
\label{tab:Btor}
\small
\begin{tabular}{c c c c c c c c c c c}
\hline\hline
$\rho_c$ (g/cc)&$M$ $(M_{\odot})$ & $R_E$ (km) & $R_P/R_E$ & $B_{max}$ (G) & $\nu$ (Hz) & ME/GE & KE/GE & $|\epsilon|$ & $h_0~(d=10~kpc)$ \\
\hline
$10^{15}$ & 1.97 & 13.97 & 0.92 & $4.5\times 10^{17}$ & 500 & $5.6\times 10^{-2}$ & $2.3\times 10^{-2}$ & 0.11 & $7.3\times 10^{-21}$ \\

$10^{15}$ & 1.91 & 12.48 & 0.986 & $2.8\times 10^{17}$ & 200 & $2\times 10^{-2}$ & $3\times 10^{-3}$ & 0.11 & $3.7\times 10^{-22}$ \\

$10^{15}$ & 1.908 & 11.99 & 0.99 & $9\times 10^{16}$ & 200 & $2\times 10^{-3}$ & $3\times 10^{-3}$ & $4\times 10^{-3}$ & $3.5\times 10^{-23}$ \\

$10^{15}$ & 2.039 & 13.31 & 0.81 & $9\times 10^{16}$ & 700 & $2.1\times 10^{-3}$ & $4.2\times 10^{-2}$ & $4\times 10^{-3}$ & $4.5\times 10^{-22}$ \\
\hline 

$5\times10^{14}$ & 1.64 & 15.8 & 0.97 & $1.5\times 10^{17}$ & 100 & $2\times 10^{-2}$ & $2.3\times 10^{-3}$ & 0.039 & $1.7\times 10^{-22}$ \\

$5\times10^{14}$ & 1.638 & 15.3 & 0.99 & $4.5\times 10^{16}$ & 100 & $1.5\times 10^{-3}$ & $1.6\times 10^{-3}$ & $3.9\times 10^{-3}$ & $1.6\times 10^{-23}$ \\ 

$5\times10^{14}$ & 1.68 & 15.63 & 0.94 & $4.5\times 10^{16}$ & 300 & $2.8\times 10^{-4}$ & $1.5\times 10^{-2}$ & $3.9\times 10^{-3}$ & $1.5\times 10^{-22}$ \\
\hline
\end{tabular}
\end{center}
\end{table*}

\subsection{Neutron stars with purely poloidal magnetic field}
\label{sec:Bpol}
\begin{figure}
\begin{center}
\includegraphics[width=\columnwidth]{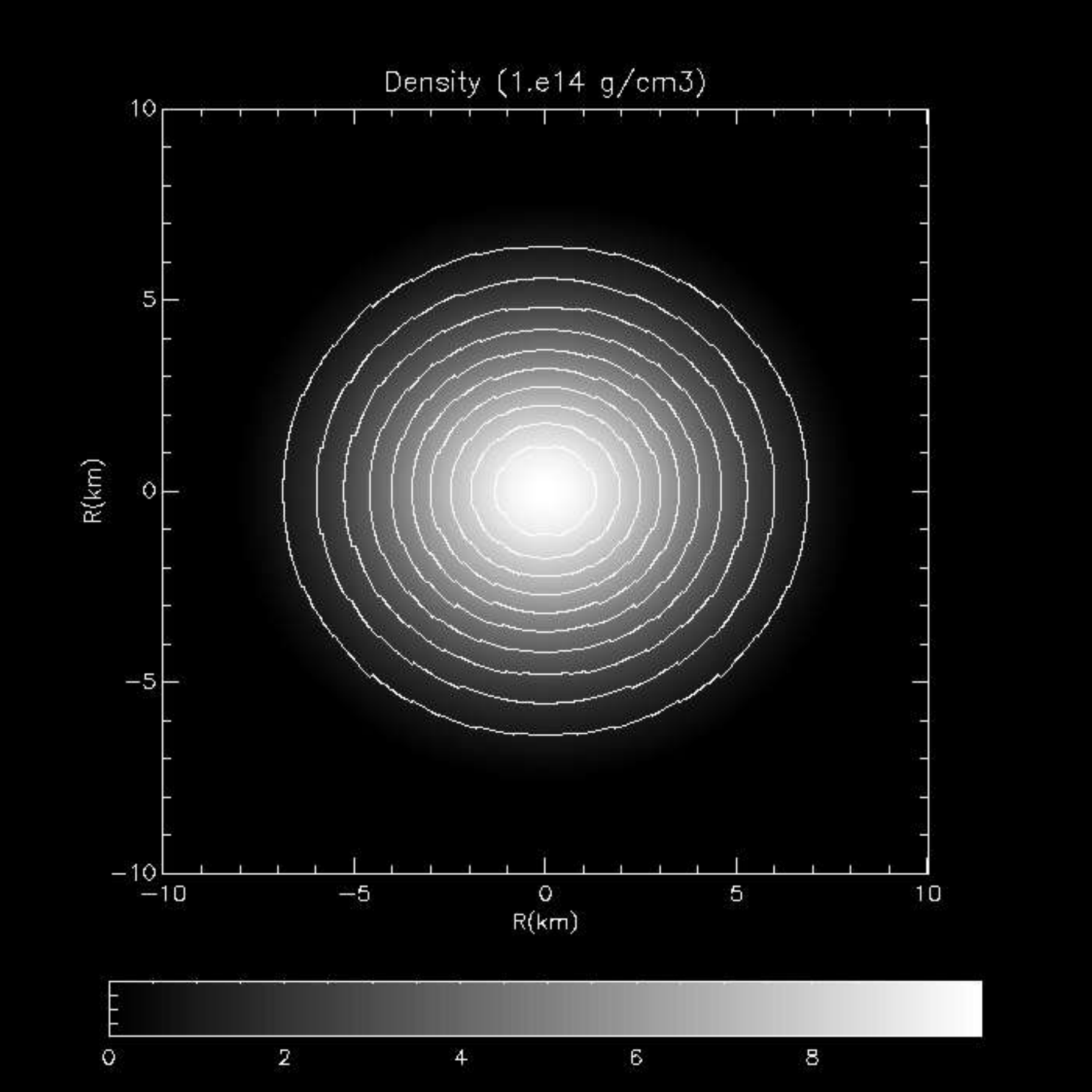}
\caption{Density isocontours of uniformly rotating poloidally magnetized NS of mass $M=1.92M_\odot$ with $\nu=50$ Hz, $B_{max}=4.4\times 10^{17}$ G, ME/GE$=2\times 10^{-2}$ and KE/GE$=2\times 10^{-4}$.}
\label{fig:polrot}
\end{center}
\end{figure}

A similar exploration is carried out for uniformly rotating NSs with a purely poloidal magnetic field. Purely poloidal magnetic field and rotation both affect the star similarly, deforming it into an oblate shape, as shown in Fig. \ref{fig:polrot}. Table \ref{tab:Bpol} shows $h_0$ for various NSs,
similarly as Table \ref{tab:Btor}.
Note that $\nu$-s in the model NSs considered in this work and presented in Tables \ref{tab:Btor} and \ref{tab:Bpol} are of the order of millisecond, which is smaller than those for
soft-gamma repeaters and anomalous X-ray pulsars (magnetars).

\begin{table*}
\begin{center}
\caption{Uniformly rotating NS with poloidal magnetic field and $\chi=3^\circ$. Here $B_{max}$ is the maximum magnetic field at the centre, when the surface field could be much smaller.}
\label{tab:Bpol}
\small
\begin{tabular}{c c c c c c c c c c c}
\hline\hline
$\rho_c$ (g/cc)&$M$ $(M_{\odot})$&$R_E$ (km) & $R_P/R_E$ & $B_{max}$ (G) & $\nu$ (Hz) & ME/GE & KE/GE & $|\epsilon|$ & $h_0~(d=10~kpc)$ \\
\hline 
$10^{15}$ & 2.045 & 12.82 & 0.81 & $4.4\times 10^{17}$ & 650 & $2.2\times 10^{-2}$ & $3.7\times 10^{-2}$ & 0.058 & $5.1\times 10^{-21}$ \\

$10^{15}$ & 1.93 & 11.99 & 0.946 & $4.4\times 10^{17}$ & 200 & $2\times 10^{-2}$ & $3\times 10^{-3}$ & 0.058 & $4.7\times 10^{-22}$ \\

$10^{15}$ & 1.921 & 11.99 & 0.945 & $4.4\times 10^{17}$ & 50 & $2\times 10^{-2}$ & $2\times 10^{-4}$ & 0.058 & $3\times 10^{-23}$ \\

$10^{15}$ & 2.017 & 12.98 & 0.85 & $9\times 10^{16}$ & 650 & $9\times 10^{-4}$ & $3.7\times 10^{-2}$ & $2.2\times 10^{-3}$ & $2.1\times 10^{-22}$ \\
\hline 

$5\times 10^{14}$ & 1.69 & 15.3 & 0.91 & $3\times 10^{17}$ & 100 & $3\times 10^{-2}$ & $1.6\times 10^{-3}$ & $9\times 10^{-2}$ & $4\times 10^{-22}$ \\

$5\times10^{14}$ & 1.64 & 15.2 & 0.98 & $4.5\times 10^{16}$ & 100 & $5.7\times 10^{-4}$ & $1.6\times 10^{-3}$ & $2.1\times 10^{-3}$ & $9\times 10^{-24}$ \\  

$5\times10^{14}$ & 1.68 & 15.5 & 0.94 & $4.5\times 10^{16}$ & 300 & $6.1\times 10^{-4}$ & $1.5\times 10^{-3}$ & $2.1\times 10^{-3}$ & $8\times 10^{-23}$ \\
\hline
\end{tabular}
\end{center}
\end{table*}

\begin{figure}
\begin{center}
\includegraphics[width=\columnwidth]{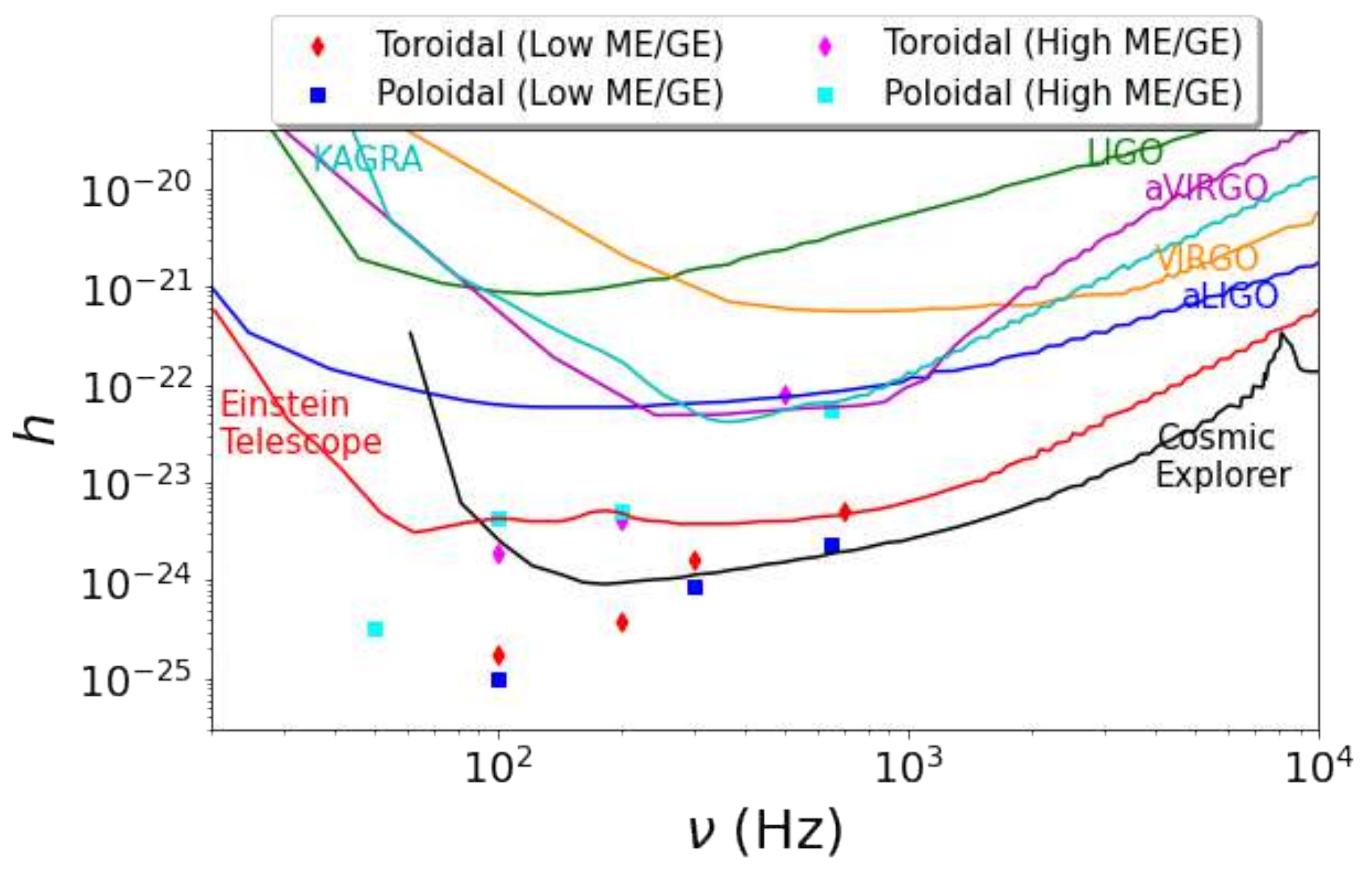}
\caption{Dimensionless GW amplitude for NSs as a function of frequency, as given in Tables \ref{tab:Btor}-\ref{tab:Bpol}, along with the sensitivity curves of various detectors. Here $h=0.0110297h_0$ with $\chi=3^\circ$.}
\label{fig:gwtorpol}
\end{center}
\end{figure}

\subsection{Neutron stars with different polytropic index}
\label{sec:EoS}
{We choose different $\Gamma$ with the same toroidal magnetic field in the core to see how $M$ and $h_0$ change with EoS. 
From Table \ref{tab:EoS}, it is clear that a more massive NS is achievable with softer EoSs, however steeper EoSs lead to more compact stars. Nevertheless, the effect of EoS on ellipticity is very small.}

\begin{table*}
\begin{center}
\caption{Uniformly rotating toroidally magnetized NS with varying polytropic index $\Gamma$ and $\chi=3^\circ$. Here $B_{max}$ is the maximum magnetic field when the surface field could be much smaller.}
\label{tab:EoS}
\small
\begin{tabular}{c c c c c c c c c c}
\hline\hline
$\Gamma$ & $\rho_c$ (g/cc)& $M$ $(M_{\odot})$&$R_E$ (km) & $R_P/R_E$ & $B_{max}$ (G) & $\nu$ (Hz) & ME/GE & KE/GE & $h_0~(d=10~kpc)$ \\
\hline
2.0 & $10^{15}$ & 1.62 & 12.156 & 0.823 & $6.1\times 10^{16}$ & 700 & $1\times 10^{-3}$ & $4.3\times 10^{-2}$ & $2.8\times 10^{-22}$ \\ 
1.95 & $10^{15}$ & 2.039 & 13.31 & 0.814 & $6\times 10^{16}$ & 700 & $8.7\times 10^{-4}$ & $4.4\times 10^{-2}$ & $2.6\times 10^{-22}$ \\ 
1.9 & $10^{15}$ & 2.457 & 14.64 & 0.8 & $5.9\times 10^{16}$ & 700 & $7\times 10^{-4}$ & $4.5\times 10^{-2}$ & $2.2\times 10^{-22}$ \\ 
\hline
\end{tabular}
\end{center}
\end{table*}

\begin{figure}
\begin{center}
\includegraphics[width=\columnwidth]{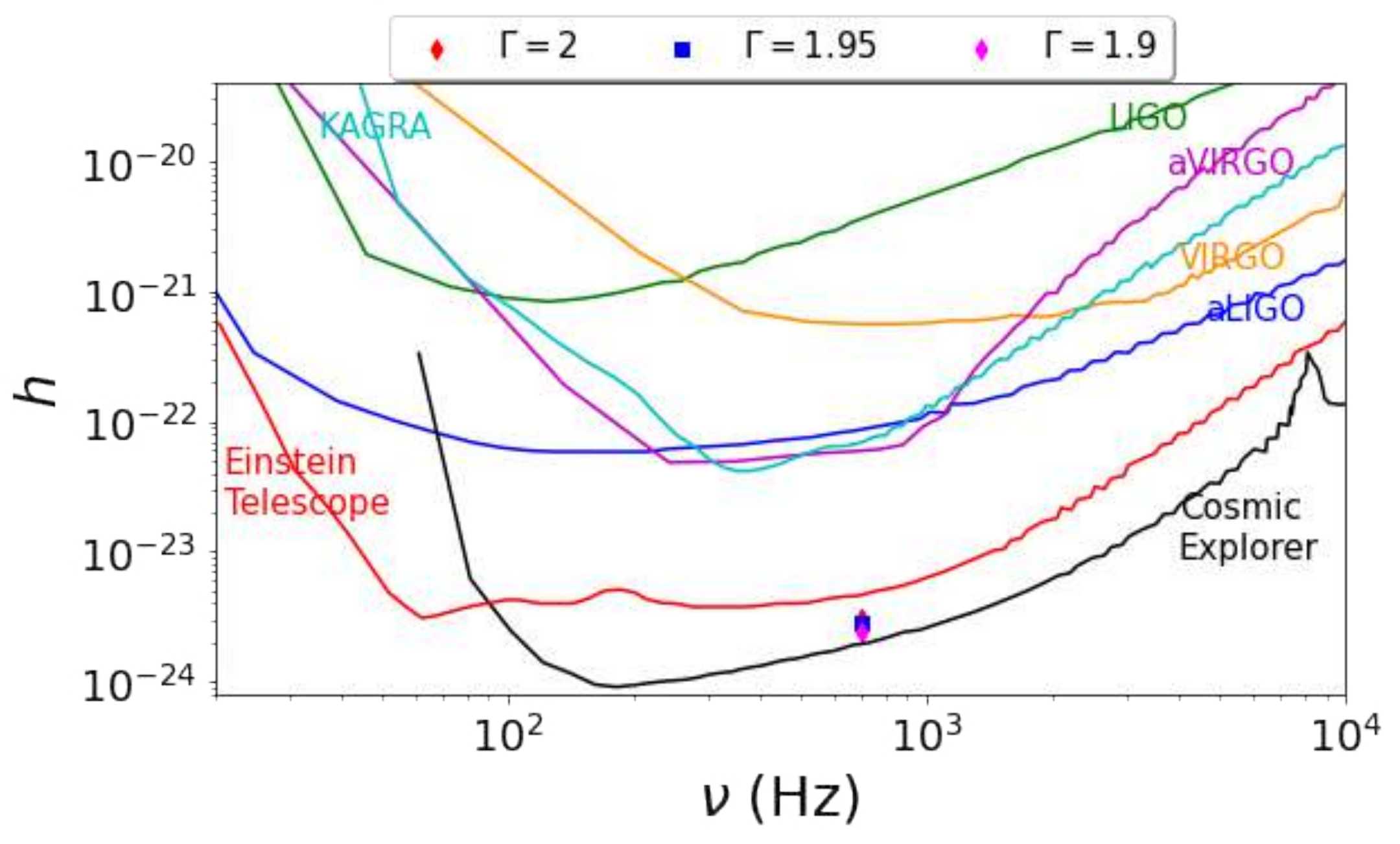}
\caption{Dimensionless GW amplitude for NSs as a function of frequency, as given in Table \ref{tab:EoS}, along with the sensitivity curves of various detectors. Here $h=0.0110297h_0$ with $\chi=3^\circ$.}
\label{fig:gwEoS}
\end{center}
\end{figure}

All the values of $h_0$ presented in Tables \ref{tab:Btor}-\ref{tab:EoS} are displayed in Figs. \ref{fig:gwtorpol} and \ref{fig:gwEoS}, along with the various sensitivity curves of different detectors. From equation (\ref{eq:gwstrain}), we know that if $\chi$ is larger, the gravitational radiation will be more efficient. Interestingly there has been no detection of CGW from NSs in LIGO, VIRGO, aLIGO, and aVIRGO so far \citep{AB2019J,AB2019JJ,P2020}. Hence, we can interpret that they are so hard to detect because the GW strain decays due to, e.g., magnetic field decay and spin-down (will be discussed in Sections \ref{sec:Bdecay} and \ref{sec:omchaidecay}). If any of them is detected in the future by Einstein Telescope, Cosmic Explorer, etc., depending on their distance from the Earth, 
its spin, magnetic field, as well as the nuclear and particle constituents of inner structure from EoS parameters, can be predicted. However, from the non-detection of CGW from NSs, we can obtain the maximum deformation supported by the NS, i.e., the upper limit of ellipticity and thus the maximum magnetic field sustained by it \citep{AB2019J,MAG2020,DP2021}.

\section{Magnetic field decay}
\label{sec:Bdecay}
In a strongly magnetized NS, Ohmic decay, ambipolar diffusion, and Hall drift are the origin of the magnetic energy release. Each of these processes may dominate the evolution depending on the magnetic field strength {and the density} of the NS. However, even though the magnetic fields throughout the NS decay by all three mechanisms, dissipation due to ambipolar diffusion may
dominate given its plausible higher field.

The timescales for Ohmic, ambipolar, and Hall effects, given by \citet{GR1992}, are, respectively,

\begin{equation}
t_{ohmic} \sim 2 \times 10^{11} \frac{L_5^2}{T_8^2} \left(\frac{\rho}{\rho_{nuc}} \right)^3 \text{ yr},
\end{equation}
\begin{equation}
t_{ambipolar} \sim \frac{5 \times 10^{15}}{T_8^6 B_{12}^2}\text{yr} + t_{ambipolar}^s,
\end{equation}
where
\begin{equation}
t_{ambipolar}^s \sim 3 \times 10^{9} \frac{L_5^2 T_8^2}{B_{12}^2} \text{  yr},
\end{equation}
and
\begin{equation}
t_{Hall} \sim 5 \times 10^{8} \frac{L_5^2}{B_{12}} \left(\frac{\rho}{\rho_{nuc}} \right) \text{ yr},
\end{equation}
where $L_5$ is the characteristic length scale of the flux loops through the outer core in units of $10^5$ cm, $T_8$ is the core temperature in units of $10^8$ K, and $B_{12}$ is the magnetic field strength in units $10^{12}$ G. We consider the core temperature as $10^9$K. Ohmic decay dominates in the weak field limit ($B\lesssim10^{11}$ G), fields of medium strength ($B\sim10^{12}-10^{13}$ G) is dissipated via Hall drift, and very strong fields ($B\gtrsim10^{14}$ G) strongly decay by ambipolar diffusion. For details on the operation of the decay mechanism, see \citealt{GR1992,HK1998}.

The magnetic field decay in NSs can be studied, following \citet{HK1998}, by solving the decay equation,

\begin{equation}
\frac{dB}{dt}=-B \left( \frac{1}{t_{ohmic}}+\frac{1}{t_{ambipolar}}+\frac{1}{t_{Hall}} \right).
\label{eq:Bdecay}
\end{equation}
By solving equation (\ref{eq:Bdecay}) through the star, we obtain 
the decay of the magnetic field profile throughout NS. The initial magnetic field profile is taken from the {\it XNS} output data.

It is evident from the density and field-dependent expressions for the above-mentioned timescales that although Hall and Ohmic decays can occur throughout the star, in the core ambipolar diffusion dominates over them \citep{GR1992} while the latter only can occur in core. Thus the magnetic field decay given by equation (\ref{eq:Bdecay}) in the core will be governed by ambipolar diffusion.

However, the ambipolar diffusion, which can be divided into solenoidal and irrotational components, may be suppressed due to the appearance of hyperons above certain density and superconductivity below certain temperature due to NS cooling \citep{BPP1969,Colpi_2000,GJS2011}. In fact, due to proton superconductivity and neutron superfluidity effects in the core, even the Hall and Ohmic decays will not occur. Thus, it has been suggested in this case that the magnetic field is evolved only in the crust by Hall drift and Ohmic dissipations, and the possible flux expulsion from the core to crust due to the interaction between neutron vortices and magnetic fluxtubes takes place. \citep{Jones1988, YAK1980,PMG2009, GJS2011}. Nevertheless, a field $> 5\times 10^{16}$ G would void superconducting activity \citep{Tilley} and the field in $10^{15}-5\times 10^{16}$ G \citep{SS2015} would make superconducting effect weak. Therefore, the field decay might remain to be unaffected or weakly affected \citep{Dall2009} therein.

\subsection{Neutron stars with purely toroidal magnetic field}
\label{sec:Bdecaytor}

\begin{figure}
\begin{center}
\includegraphics[width=\columnwidth]{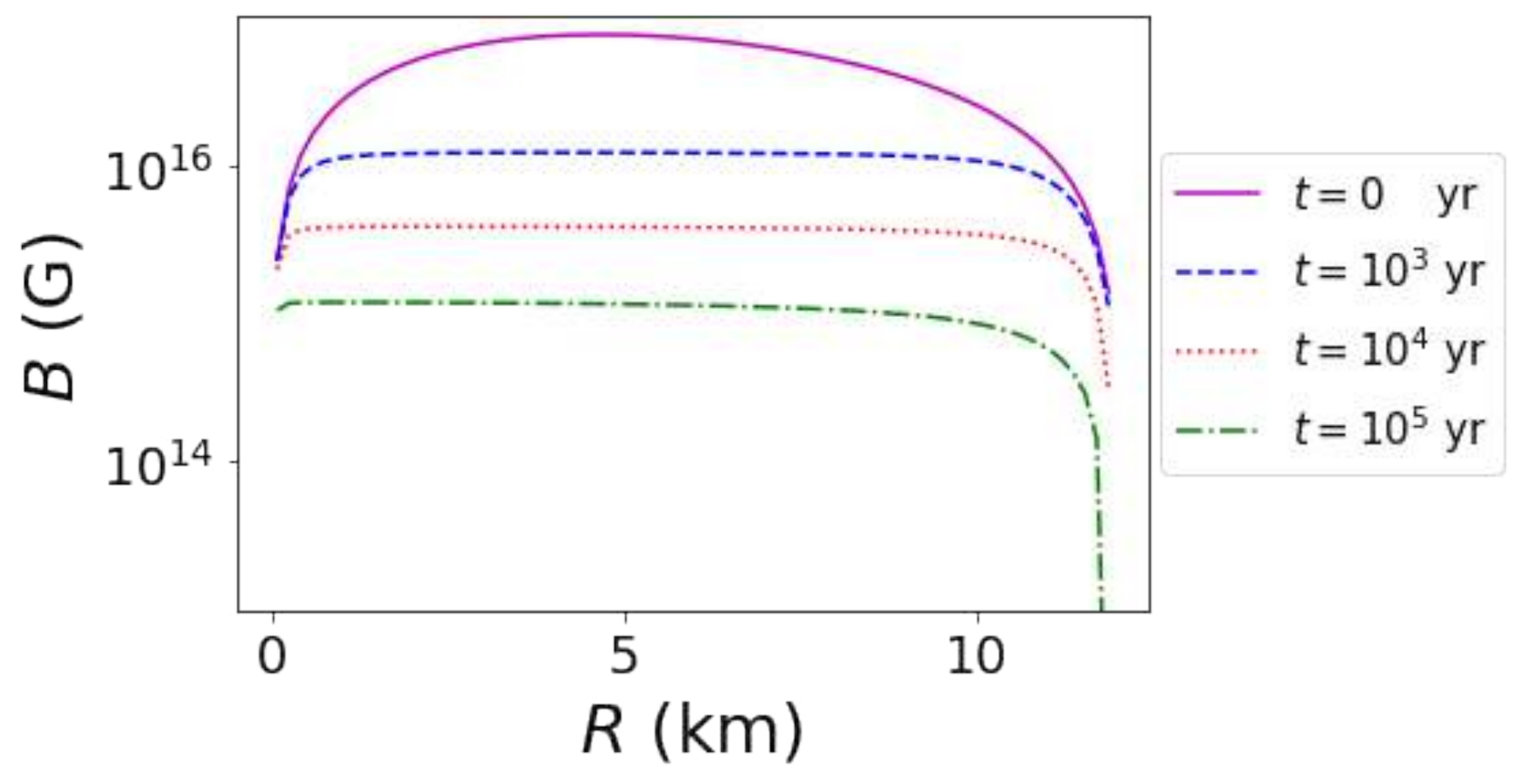}
\caption{Magnetic field as a function of radius, before and after magnetic field decay, as given in Table \ref{tab:Bdecaytor} for toroidal field.}
\label{fig:Btorprof}
\end{center}
\end{figure}

\begin{figure}
\begin{center}
\includegraphics[width=\columnwidth]{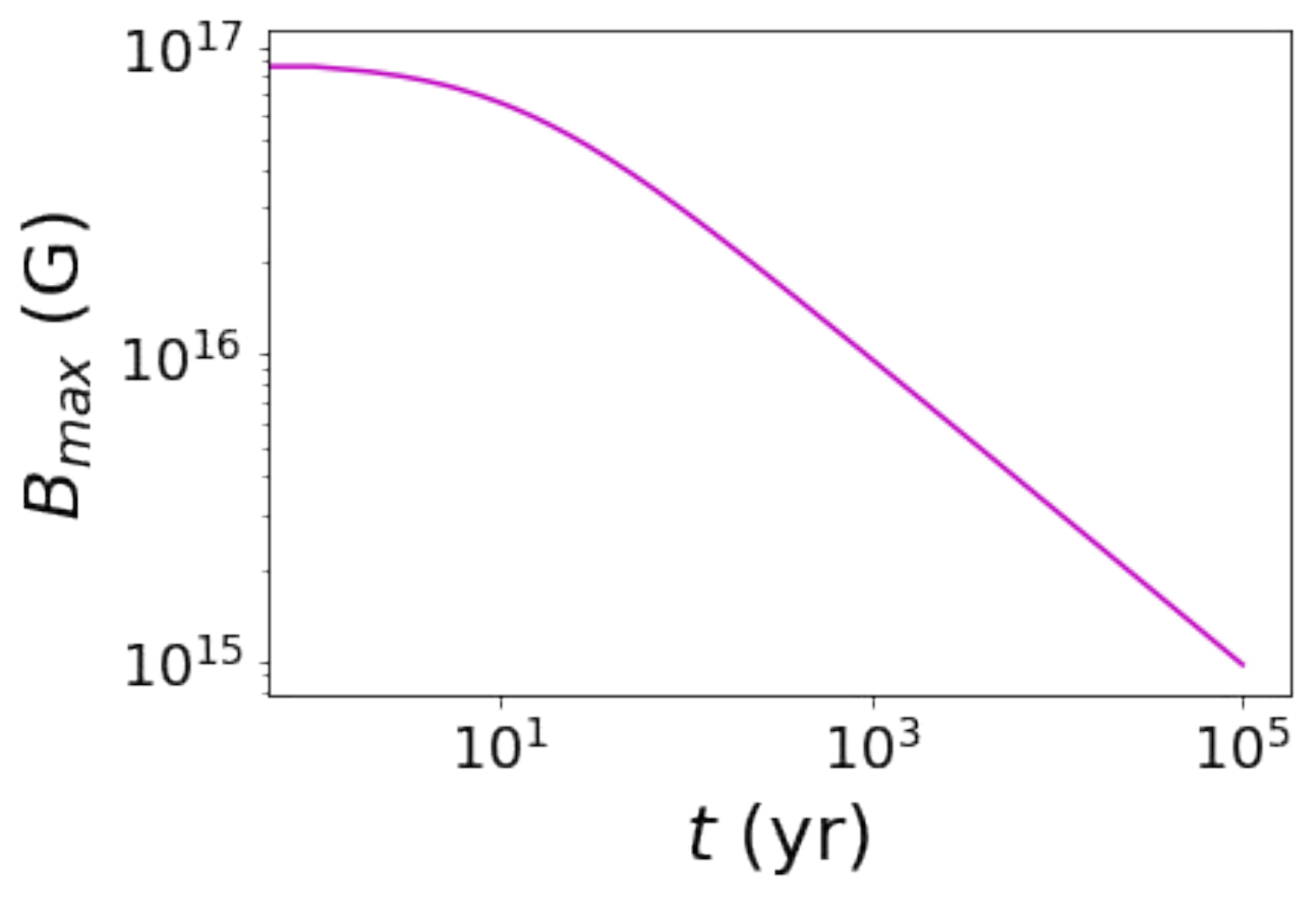}
\caption{Maximum magnetic field as a function of time of a NS, as given in Table \ref{tab:Bdecaytor} for toroidal field.}
\label{fig:Btordecay}
\end{center}
\end{figure}

\begin{figure}
\begin{center}
\includegraphics[width=\columnwidth]{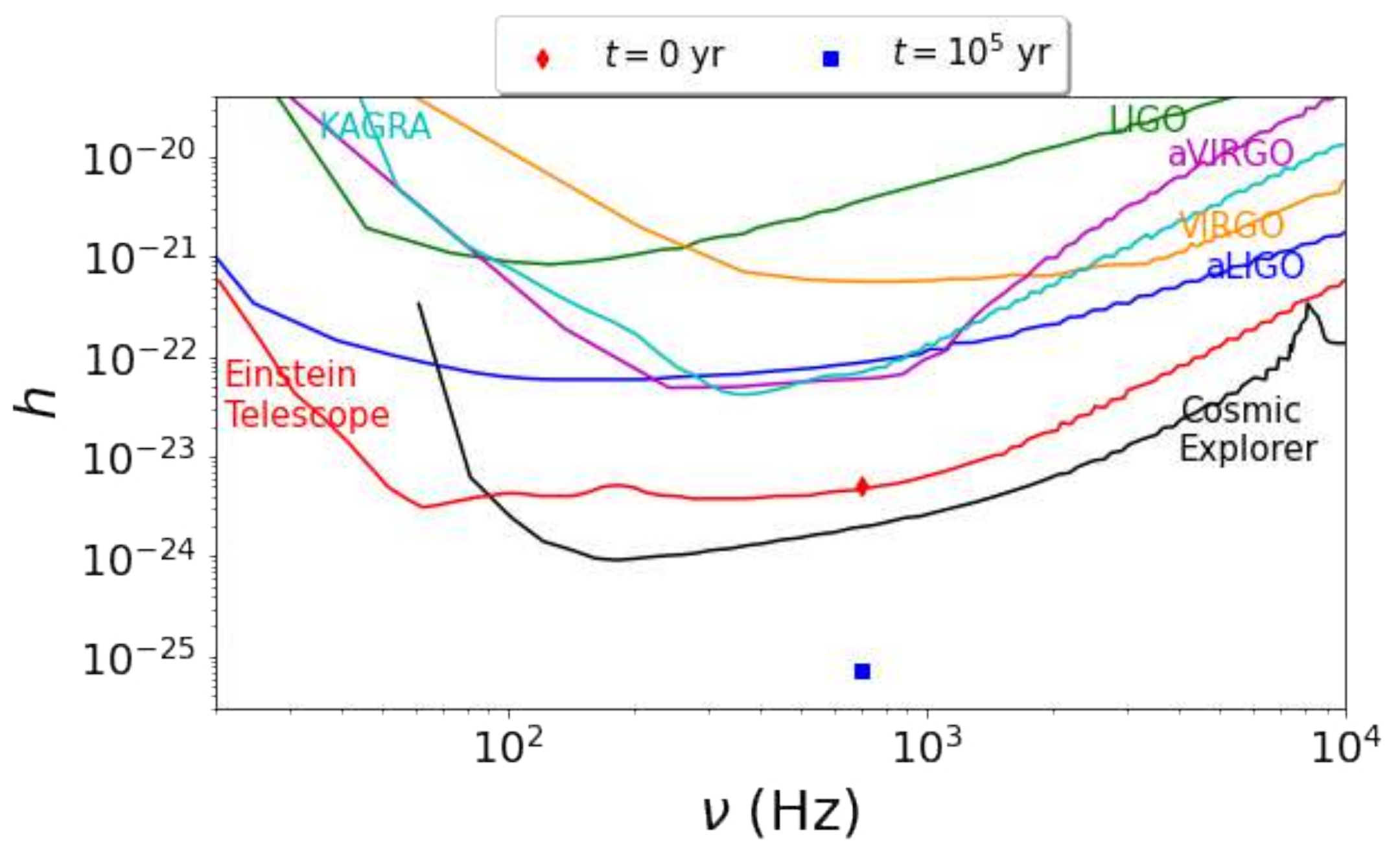}
\caption{Dimensionless GW amplitude for NSs before and after magnetic field decay, as given in Table \ref{tab:Bdecaytor} for toroidal fields. Here $h=0.0110297h_0$ with $\chi=3^\circ$.}
\label{fig:Btordecaygw}
\end{center}
\end{figure}

We start with an initial uniformly rotating NS model based on {\it XNS} with $\Gamma=1.95$ and $\rho_c=10^{15}$ g cm\textsuperscript{-3}. Substituting {\it XNS} output in equation
(\ref{eq:Bdecay}) as the initial condition, we obtain the field profile 
as a function of time. Fig. \ref{fig:Btorprof} shows the magnetic field
profiles in a NS at different ages. For the maximum magnetic field of the star, 
Fig. \ref{fig:Btordecay} shows how the field decays with time. After certain 
time, we can estimate $B_{max}$ from Fig. \ref{fig:Btordecay} and can model 
another NS with such a $B_{max}$. Fig. \ref{fig:Btordecaygw} shows how 
the GW amplitude decays with time as $B$ decays, assuming $\Omega$ and $\chi$
fixed. As seen in the figure, initially, NS could be detected by the 
Einstein Telescope and Cosmic Explorer, however, after a certain time, it will 
be undetectable. In Fig. \ref{fig:Btorprof}, all the magnetic field profiles 
are shown, assuming the same radius for the NS. Nevertheless, with decreasing $B$, the radius 
of NS should decrease because the outward magnetic pressure decreases. 
However, for the present case, the change of radius is very negligible 
(one can look at Table \ref{tab:Bdecaytor}), hence the assumption is valid.

If the superconducting effects suppress the ambipolar (and other decay) effect, i.e., if the decay due to ambipolar diffusion does not work, the field decay in the core would be suppressed. This allows the star to emit GW for a longer time until the Meissner effect expels the flux from the core to the crust. 

Nevertheless, note that with time, $\Omega$ and $\chi$ 
may decrease significantly. Hence, the GW amplitude at $t>0$ presented in Table 
\ref{tab:Bdecaytor} and Fig. \ref{fig:Btordecaygw} need not be accurate.
The decays of $\Omega$ and $\chi$  and their effects on GW amplitude will be discussed in detail in Section \ref{sec:omchaidecay}.
  
\begin{table*}
\caption{Uniformly rotating NS with the toroidal magnetic field at different times after birth, where $\rho_c=10^{15}$ g cm$^{-3}$.}
\label{tab:Bdecaytor}
\begin{center}
\begin{tabular}{c c c c c c c c c c c} 
\hline\hline
$t$(yr)&$M$ $(M_{\odot})$&$R_E$ (km) & $R_P/R_E$ & $B_{max}$ (G) & $\nu$ (Hz) & ME/GE & KE/GE & $|\epsilon|$ & $h_0~(d=10~kpc)$ \\
\hline
0 & 2.039 & 13.31 & 0.81 & $9\times 10^{16}$ & 700 & $2.1\times 10^{-3}$ & $4.2\times 10^{-2}$ & $5.1\times 10^{-4}$ & $4.5\times 10^{-22}$ \\

$10^{5}$ & 2.039 & 13.15 & 0.82 & $10^{15}$ & 700 & $2\times 10^{-7}$ & $4.3\times 10^{-2}$ & $6.2\times 10^{-5}$ & $6.5\times 10^{-24}$ \\
\hline
\end{tabular}
\end{center}
\end{table*}

\subsection{Neutron stars with purely poloidal magnetic field}
\label{sec:Bdecaypol}

\begin{figure}
\begin{center}
\includegraphics[width=\columnwidth]{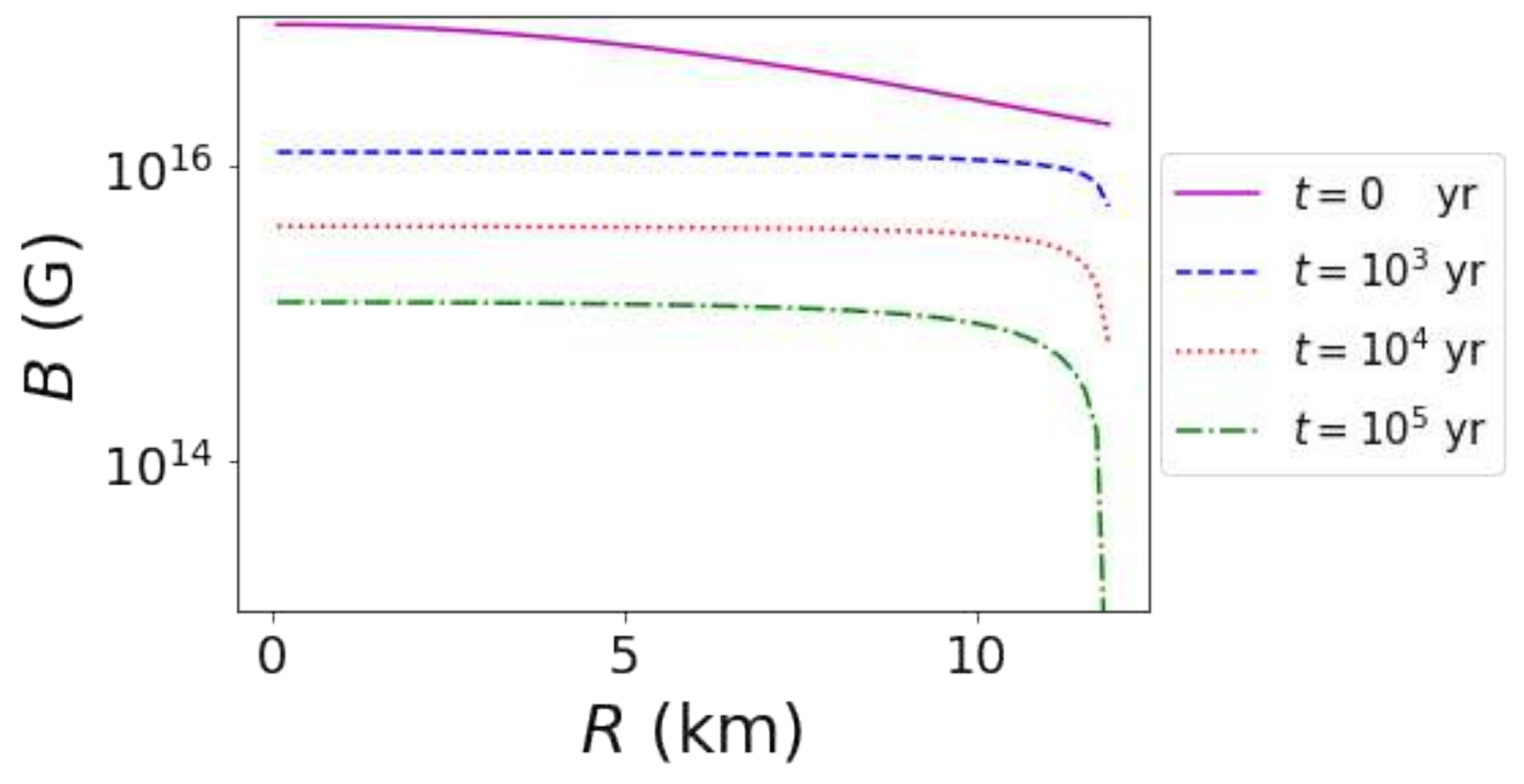}
\caption{Magnetic field as a function of radius, before and after magnetic field decay, as given in Table \ref{tab:Bdecaypol} for poloidal fields.}
\label{fig:Bpolprof}
\end{center}
\end{figure}

\begin{figure}
\begin{center}
\includegraphics[width=\columnwidth]{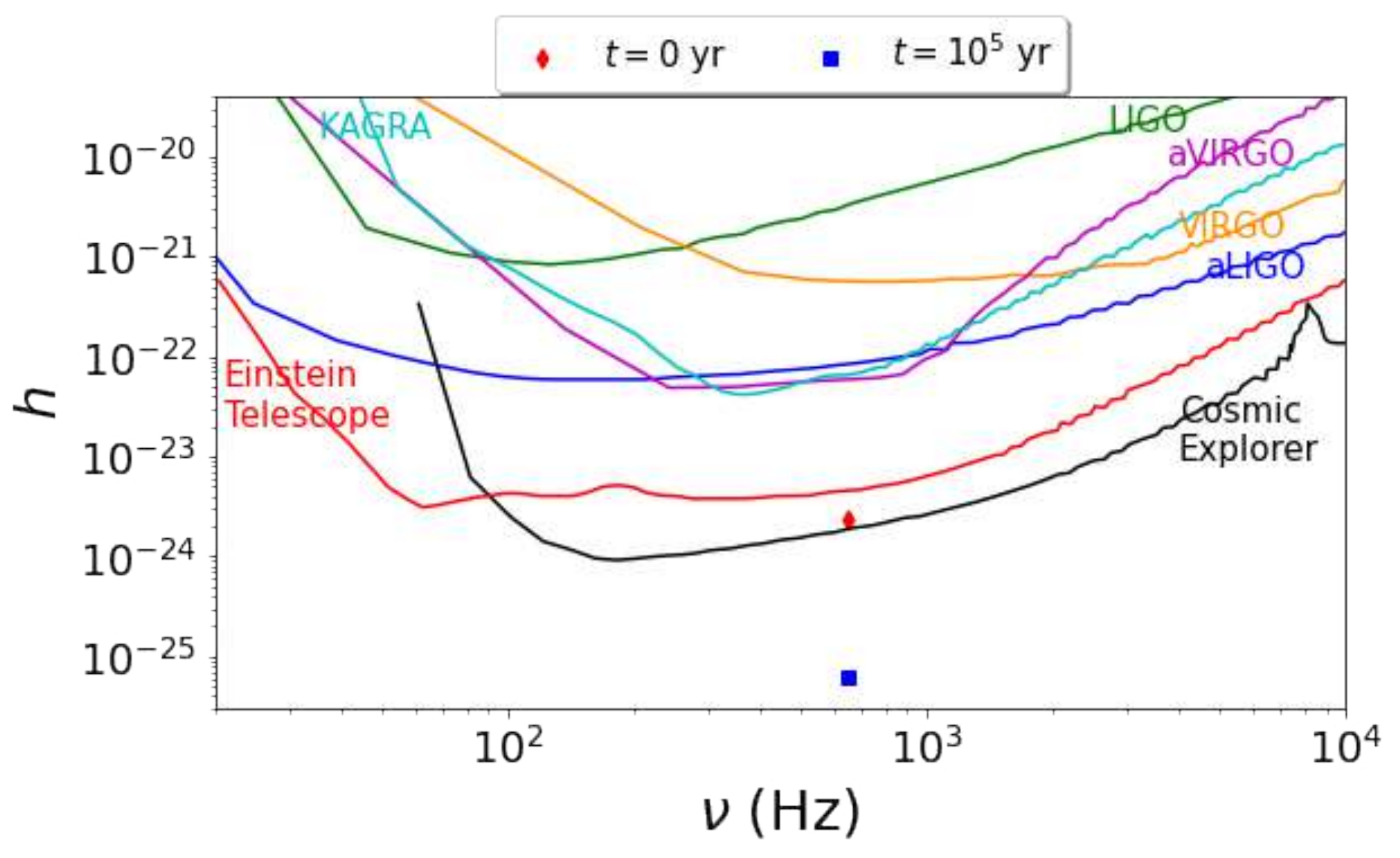}
\caption{Dimensionless GW amplitude for NSs before and after magnetic field decay, as given in Table \ref{tab:Bdecaypol} for poloidal fields. Here $h=0.0110297h_0$ with $\chi=3^\circ$.}
\label{fig:Bpoldecaygw}
\end{center}
\end{figure}

A similar exploration as in Section \ref{sec:Bdecaytor} is carried out for a uniformly rotating NS with a purely poloidal magnetic field. The magnetic field profiles after a certain age of the star are shown in Fig. \ref{fig:Bpolprof}. 
Further, Fig. \ref{fig:Bpoldecaygw} and Table \ref{tab:Bdecaypol} show that the GW amplitude decays with time as $B$ decays, provided $\Omega$ and $\chi$ are fixed (which need not necessarily be true, as mentioned above). See Section \ref{sec:omchaidecay} below.
  
\begin{table*}
\caption{Uniformly rotating NS with the poloidal magnetic field at different times after birth, where $\rho_c=10^{15}$ g cm$^{-3}$.}
\label{tab:Bdecaypol}
\begin{center}
\begin{tabular}{c c c c c c c c c c c} 
\hline\hline
$t$(yr)&$M$ $(M_{\odot})$&$R_E$ (km) & $R_P/R_E$ & $B_{max}$ (G) & $\nu$ (Hz) & ME/GE & KE/GE & $|\epsilon|$ & $h_0~(d=10~kpc)$ \\
\hline
0 & 2.017 & 12.98 & 0.847 & $9\times 10^{16}$ & 650 & $9\times 10^{-3}$ & $3.7\times 10^{-2}$ & $5.4\times 10^{-4}$ & $2.1\times 10^{-22}$ \\

$10^{5}$ & 2.016 & 12.98 & 0.847 & $10^{15}$ & 650 & $9.7\times 10^{-8}$ & $3.7\times 10^{-2}$ & $6.2\times 10^{-5}$ & $5.6\times 10^{-24}$ \\ 
\hline
\end{tabular}
\end{center}
\end{table*}

\section{Spin-down and time evolution of obliquity angle}\label{sec:omchaidecay}

A triaxial pulsating NS can simultaneously emit dipole and gravitational radiations, which are incorporated with the dipole and quadrupolar luminosities. The dipole luminosity for an axisymmetric WD was discussed by \citet{MELATOS2000}, which is applicable for NS as well, given by

\begin{equation}
L_D=\frac{B_p^2 R_p^6 \Omega^4}{2c^3} \sin^2 \chi F(x_0),
\label{eq:LD}
\end{equation}
where $x_0=R_0\Omega/c$, $B_p$ is the strength of the magnetic field at the pole, $R_p$ is poler radius, $R_0$ is the average radius of NS and the function $F(x_0)$ is defined as
\begin{equation}
F(x_0)=\frac{x_0^4}{5(x_0^6-3x_0^4+36)}+\frac{1}{3(x_0^2+1)}.
\end{equation}
Similarly, the quadrupolar GW luminosity is given by \citep{ZS1979}

\begin{equation}
L_{GW}=\frac{2G}{5c^5} (I_{zz}-I_{xx})^2 \Omega^6 \sin^2 \chi (1+15 \sin^2 \chi).
\label{eq:LGW}
\end{equation}
The rotational frequency of pulsating NS decreases over time due to the extraction of angular momentum by gravitational and electromagnetic dipole radiations, which further leads to changes in obliquity angle, $\chi$. For an oblique rotator, the evolution equations for $\Omega$ and $\chi$ (for spin-down and alignment) will be involved with a combination of GW radiation and electromagnetic energy-loss terms \citep{CH1970,KMMB2020}, given by

\begin{equation}
\begin{split}
\frac{d(\Omega I_{z^{\prime}z^{\prime}})}{dt}=-\frac{2G}{5c^5} (I_{zz}-I_{xx})^2 \Omega^5 \sin^2 \chi (1+15 \sin^2 \chi)\\
-\frac{B_p^2 R_p^6 \Omega^3}{2c^3} \sin^2 \chi F(x_0)
\end{split}
\label{eq:dwdt}
\end{equation}
and
\begin{equation}
\begin{split}
I_{z^{\prime}z^{\prime}}\frac{d\chi}{dt}=-\frac{12G}{5c^5} (I_{zz}-I_{xx})^2 \Omega^4 \sin^3 \chi \cos \chi\\
-\frac{B_p^2 R_p^6 \Omega^2}{2c^3} \sin \chi \cos \chi F(x_0).
\end{split}
\label{eq:dxdt}
\end{equation}

The set of equations (\ref{eq:dwdt}) and (\ref{eq:dxdt}) will be solved simultaneously to obtain the timescale over which a NS can radiate or behave like a pulsar. To solve equations (\ref{eq:dwdt}) and (\ref{eq:dxdt}), we need to supply the various quantities, such as $I_{xx}$, $I_{zz}$, $B_P$, and $R_P$ at the initial time, which are the output of particular NS model from the {\it XNS} code. In the following, we study the evolutions of $\Omega$ and $\chi$ for various field geometries and their consequences on CGW.

\subsection{Neutron stars with purely poloidal magnetic field}
\label{sec:wdecaypol}
We choose $\rho_c=10^{15}$ g ${\text{cm}}^{-3}$, $\Gamma=1.95$, and various $B_p$ along with the initial $\chi=3^\circ$, so that we have an idea about the timescales for purely poloidally magnetized NSs behaving as pulsars.

We can treat the NSs as oscillating/rotating dipoles; hence, the dipole luminosity formula is applicable. All the different $B_p$s are given in Table \ref{tab:polwxdecay} along with the respective $M$, $R_p$, and $h_0$ at $t=0$ assuming $d = 10$ kpc. We restrict ME/GE to less than $10^{-3}$ so that the magnetized NSs are surely stable \citep{KEH1989,BR2009}. Below we discuss the time evolutions of the rotational frequency, $\chi$, and the various luminosities of NSs.

\begin{table*}
\caption{Poloidal magnetic field with $\rho_c=10^{15}$ g ${\text{cm}}^{-3}$ and $\nu=100$ Hz. $I_{xx}$ and $I_{zz}$ are in the units $4.5\times10^{43}$ g cm\textsuperscript{2}.}
\label{tab:polwxdecay}
\begin{center}
\small
\begin{tabular}{c c c c c c c c c c c c c} 
\hline\hline
$M$&$R_P$ & $B_P$ & ME/GE & KE/GE & $I_{xx}$& $I_{zz}$& $L_{GW}$ & $L_D$ & $h_0~(d=10~kpc)$ & $T_\Omega$ & $T_\chi$\\
$(M_{\odot})$ & (km)&(G)& & & & &(erg/s) &(erg/s)& at $t=0$ &(yr) &(yr)\\
\hline

1.9 & 11.99 & $1\times 10^{15}$ & $4.7\times 10^{-6}$ & $7.7\times 10^{-4}$ & 11.41 & 11.44 & $3.5\times 10^{38}$ & $7.7\times 10^{42}$ & $4.2\times 10^{-24}$ & 0.45 & 0.003\\

1.9 & 11.99 & $1\times 10^{12}$ & $1\times 10^{-11}$ & $7.7\times 10^{-4}$ & 11.41 & 11.44 & $3.5\times 10^{38}$ & $7.7\times 10^{36}$ & $1.3\times 10^{-25}$ & 439278.4 & 3507.5\\
\hline
\end{tabular}
\end{center}
\end{table*}

\subsubsection{Case I: $L_D \gg L_{GW}$}

\begin{figure}
\begin{center}
\includegraphics[width=\columnwidth]{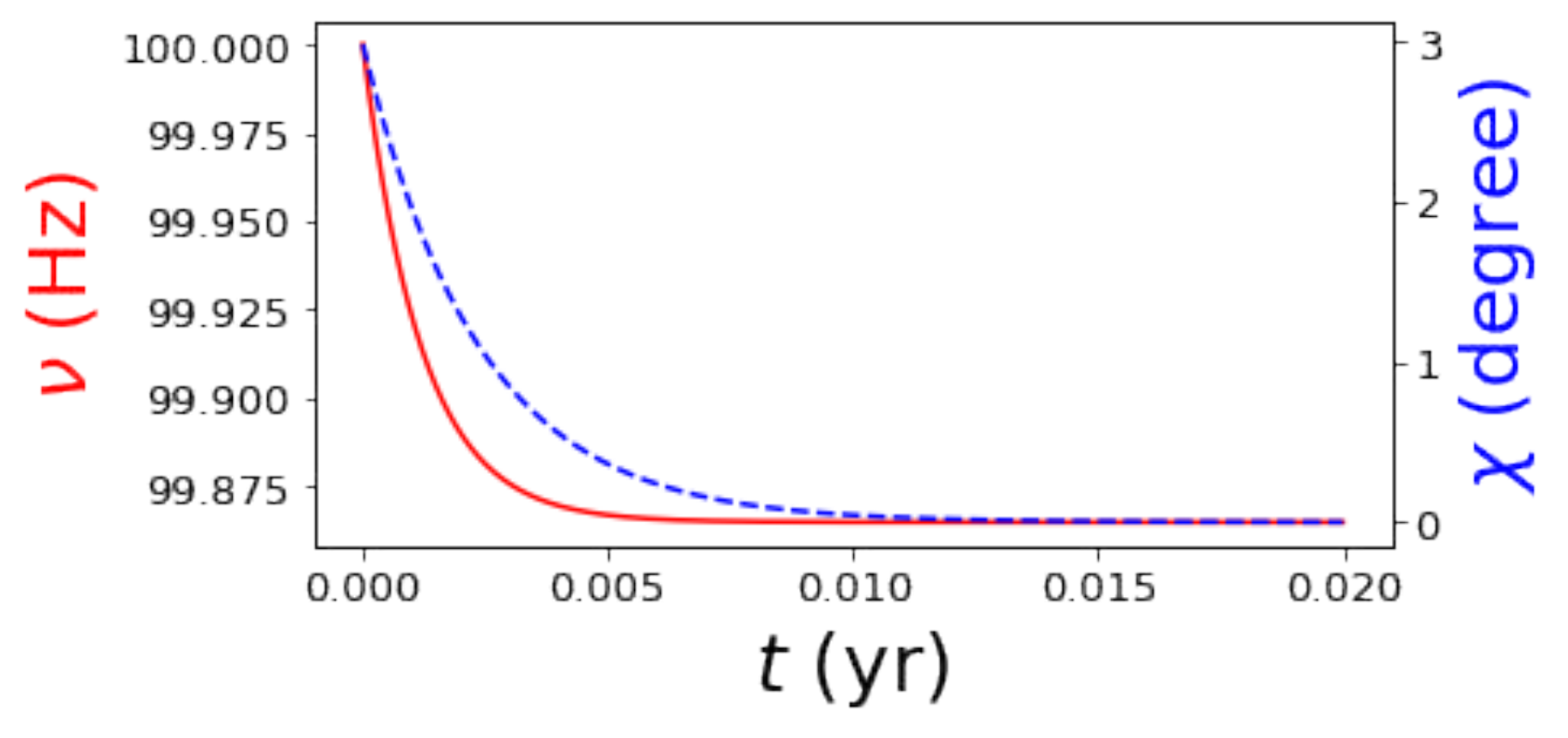}
\includegraphics[width=\columnwidth]{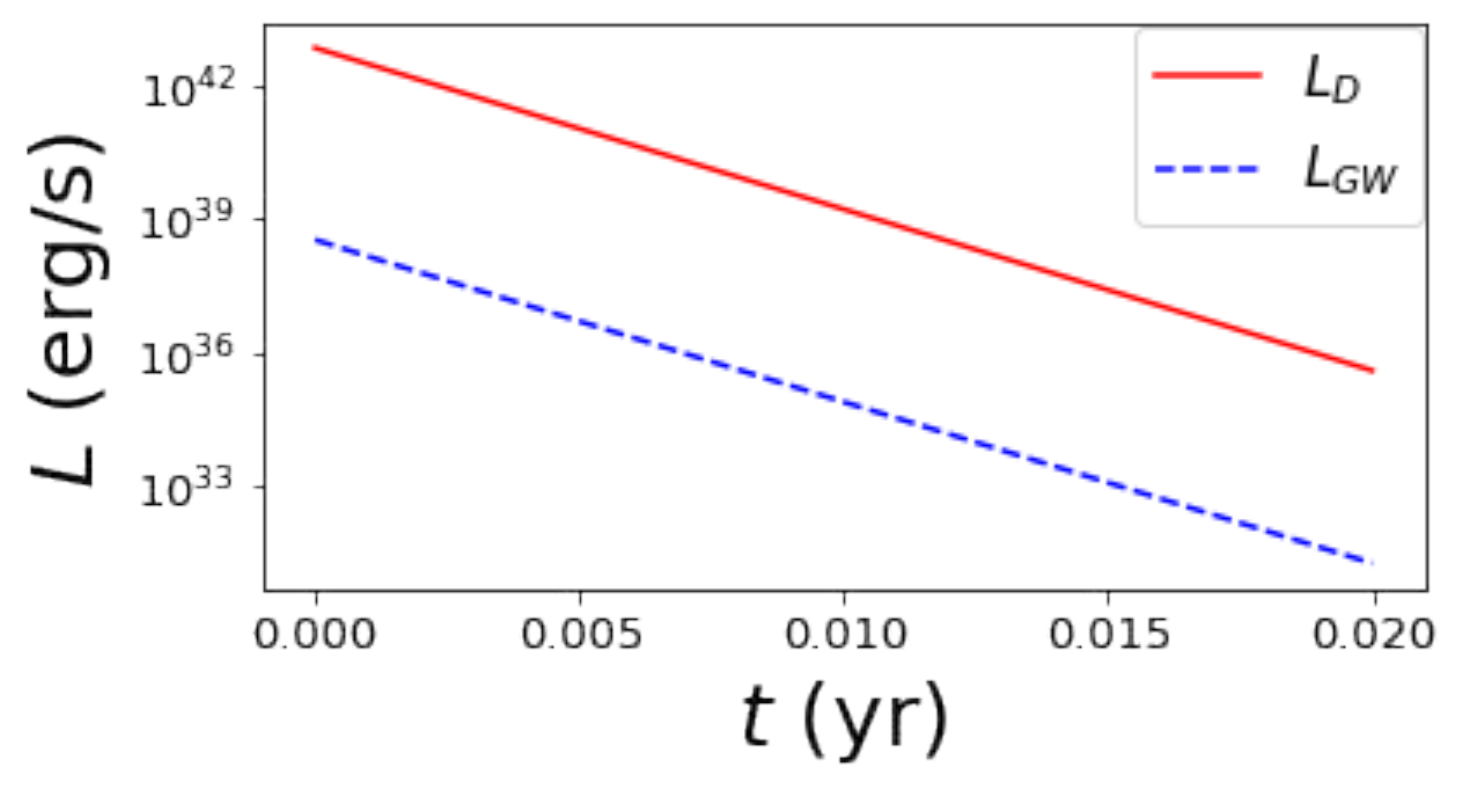}
\caption{$L_D \gg L_{GW}$:
	Variations of $\nu$, $\chi$, $L_{GW}$ and $L_D$ with time for $B_P=10^{15}$ G, initial $\nu=100$ Hz, $\chi=3^\circ$, as given in Table \ref{tab:polwxdecay}. Red (solid) and blue (dashed) lines show the variations of $\nu$ and $\chi$, respectively.}
\label{fig:polwxldlgw}
\end{center}
\end{figure}

NSs holding a high magnetic field have $L_D \gg L_{GW}$ because $L_D$ increases with the poloidal magnetic field. Thus, the luminosity is dominated by $L_D$, while the decay rates of rotation frequency and obliquity angle are proportional to total luminosity; $L_D$ governs the timescale. Furthermore, the total luminosity of the NS decreases with time due to a decrease in $\chi$ and/or $\Omega$. When $L_D \gg L_{GW}$, $\chi$ decays faster than $\Omega$. Let us denote the timescale for the change of $\chi$ from its initial value to 0 be $T_\chi$ and the corresponding $\Omega$ to saturate be $T_\Omega$. \citet{KMMB2020} calculated the timescales by integrating the equations (\ref{eq:dwdt}) and (\ref{eq:dxdt}) approximately, assuming $I_{z^{\prime}z^{\prime}}$ is not changing with time. The timescales are given by
\begin{equation}
T_\Omega \sim \left(\frac{2I_{z^{\prime}z^{\prime}} c^3}{B_p^2 R_p^6 \Omega^2 F(x_0)}\right)\frac{1}{2 \sin^2 \chi}
\end{equation}
and
\begin{equation}
	T_\chi \sim \left(\frac{2I_{z^{\prime}z^{\prime}} c^3}{B_p^2 R_p^6 \Omega^2 F(x_0)}\right) \ln (\cot \chi).
\end{equation}
In the range $0^\circ \leq \chi \leq 3^\circ$, we will have $\ln (\cot \chi) \ll 1/2 \sin^2 \chi$, which suggests $T_\chi \ll T_\Omega$. This demonstrates that $\chi$ becomes 0 very fast, and the NS starts rotating with a different angular frequency (thus linear frequency) than it originally possesses, which can be seen from Fig. \ref{fig:polwxldlgw}, along with the decays of $L_D$ and $L_{GW}$. For instance, if $M=1.9 M_\odot$, $B_P=10^{15}$ G, $R_P = 12$ km, and at $t = 0$, $\nu=100$ Hz and $\chi=3^\circ$, such that $L_D \gg L_{GW}$, then $T_\Omega \sim 30$ yr and $T_\chi \sim 0.24$ yr, hence $T_\chi \ll T_\Omega$. Due to high $L_D$, and thus very fast decay of $\chi$ and $L_D$, this NS cannot emit radiation for a long. The future GW detectors may detect such NSs just for a very short duration of time only or may not be able to detect at all.

\subsubsection{Case II: $L_{GW} \gg L_D$}

If the magnetic field is lower, the NSs may have $L_{GW} \gg L_D$. Then, luminosity decreases slowly, and the NS can radiate for a longer period. For $L_{GW} \gg L_D$, the decay timescales are obtained by integrating equations (\ref{eq:dwdt}) and (\ref{eq:dxdt}) \citep{KMMB2020}, given by

\begin{equation}
T_\Omega^\prime \sim \left(\frac{5I_{z^{\prime}z^{\prime}} c^5}{2G(I_{zz}-I_{xx})^2 \Omega^4}\right) \frac{1}{4 \sin^2 \chi (1+15 \sin^2 \chi)}
\end{equation}
and
\begin{equation}
T_\chi^\prime \sim \left(\frac{5I_{z^{\prime}z^{\prime}} c^5}{2G(I_{zz}-I_{xx})^2 \Omega^4}\right)\frac{1}{12} \left( \frac{1}{\sin^2 \chi}+ 2 \ln \cot \chi \right).
\end{equation}

\begin{figure}
\begin{center}
\includegraphics[width=\columnwidth]{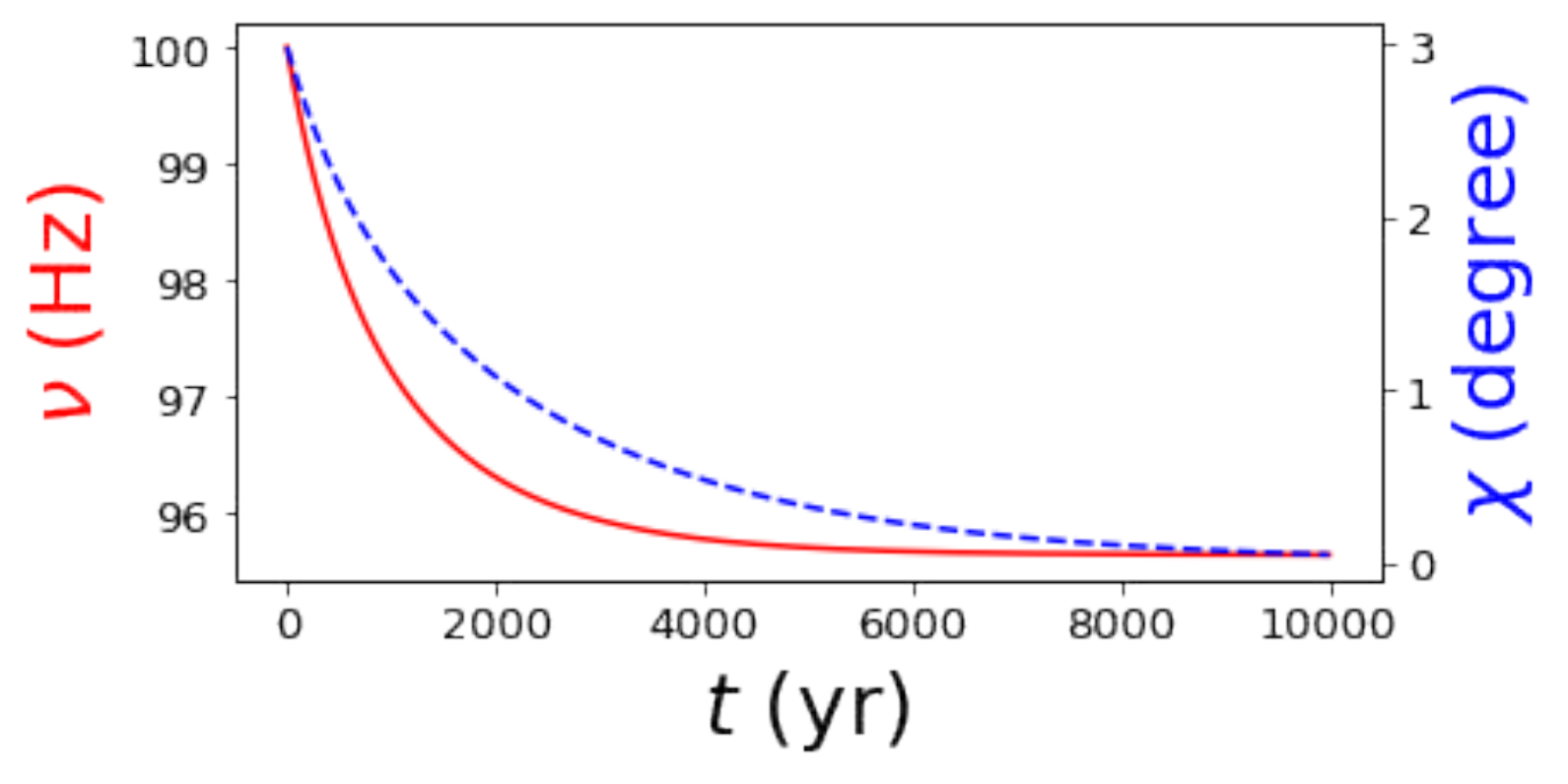}
\includegraphics[width=\columnwidth]{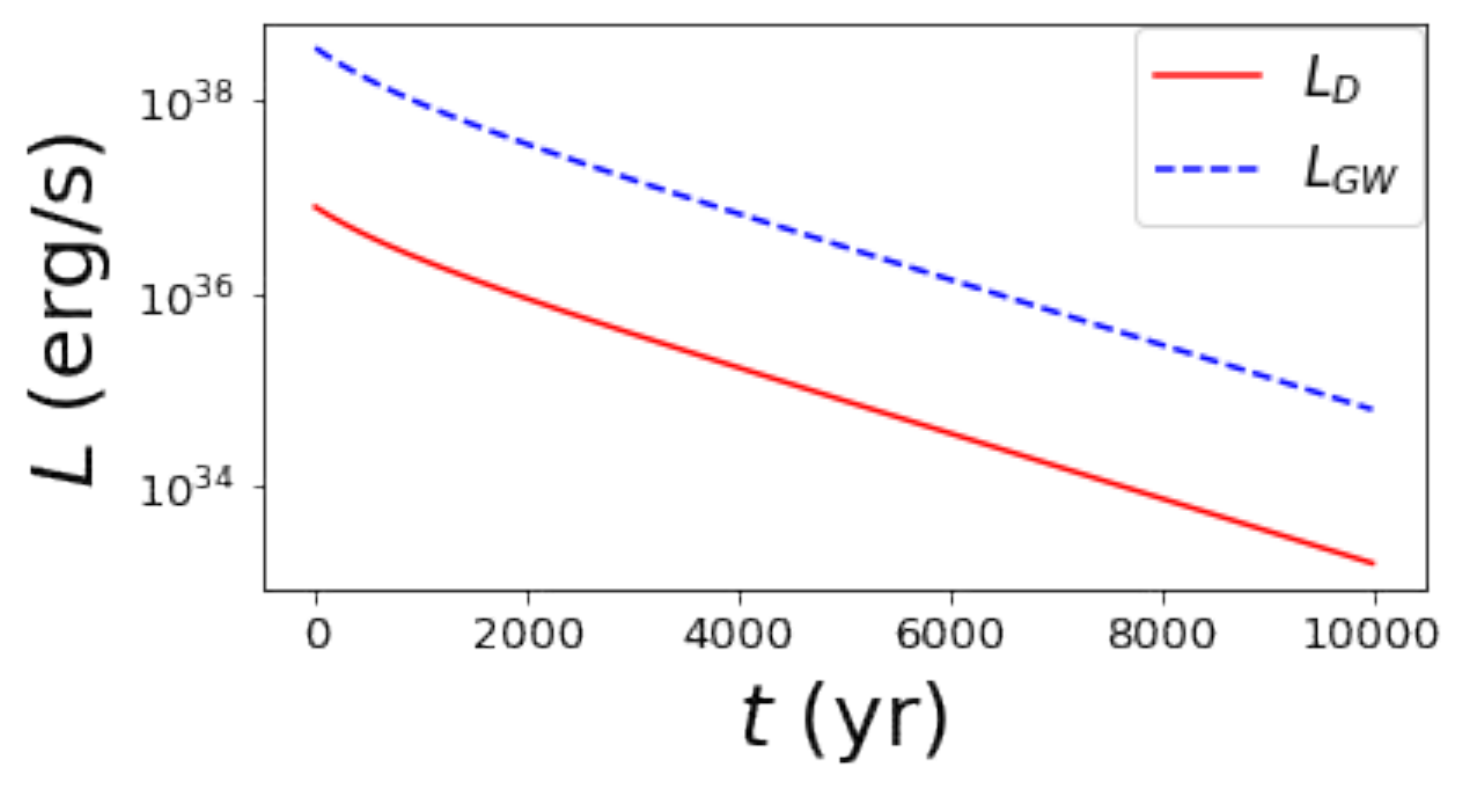}
\caption{$L_{GW} \gg L_D$:
	Variations of $\nu$, $\chi$, $L_{GW}$ and $L_D$ with time for $B_P=10^{12}$ G, initial $\nu=100$ Hz, $\chi=3^\circ$, as given in Table \ref{tab:polwxdecay}. Red (solid) and blue (dashed) lines show the variations of $\nu$ and $\chi$, respectively.}
\label{fig:polwxlgwld}
\end{center}
\end{figure}

In the range, $0^\circ \leq \chi \leq 3^\circ$, $\Omega$, and $\chi$ decay simultaneously for a long time before approaching a saturated value and zero, respectively. This also can be verified from Fig. \ref{fig:polwxlgwld}. 
This further allows $L_{GW}$ and $L_D$ to remain higher for longer. The decay of $\Omega$ and $\chi$ is governed by GW radiation mainly. If $M=1.9 M_\odot$, $B_P=10^{12}$ G, $R_P = 12$ km, and at $t = 0$, $\nu=100$ Hz and $\chi=3^\circ$, such that $L_{GW} \gg L_D$, then $T_\Omega \sim 4.4 \times 10^5$ yr and $T_\chi \sim 3.5\times10^3$ yr.

\subsection{Neutron stars with purely toroidal magnetic field}
\label{sec:wdecaytor}
\begin{figure}
\begin{center}
\includegraphics[width=\columnwidth]{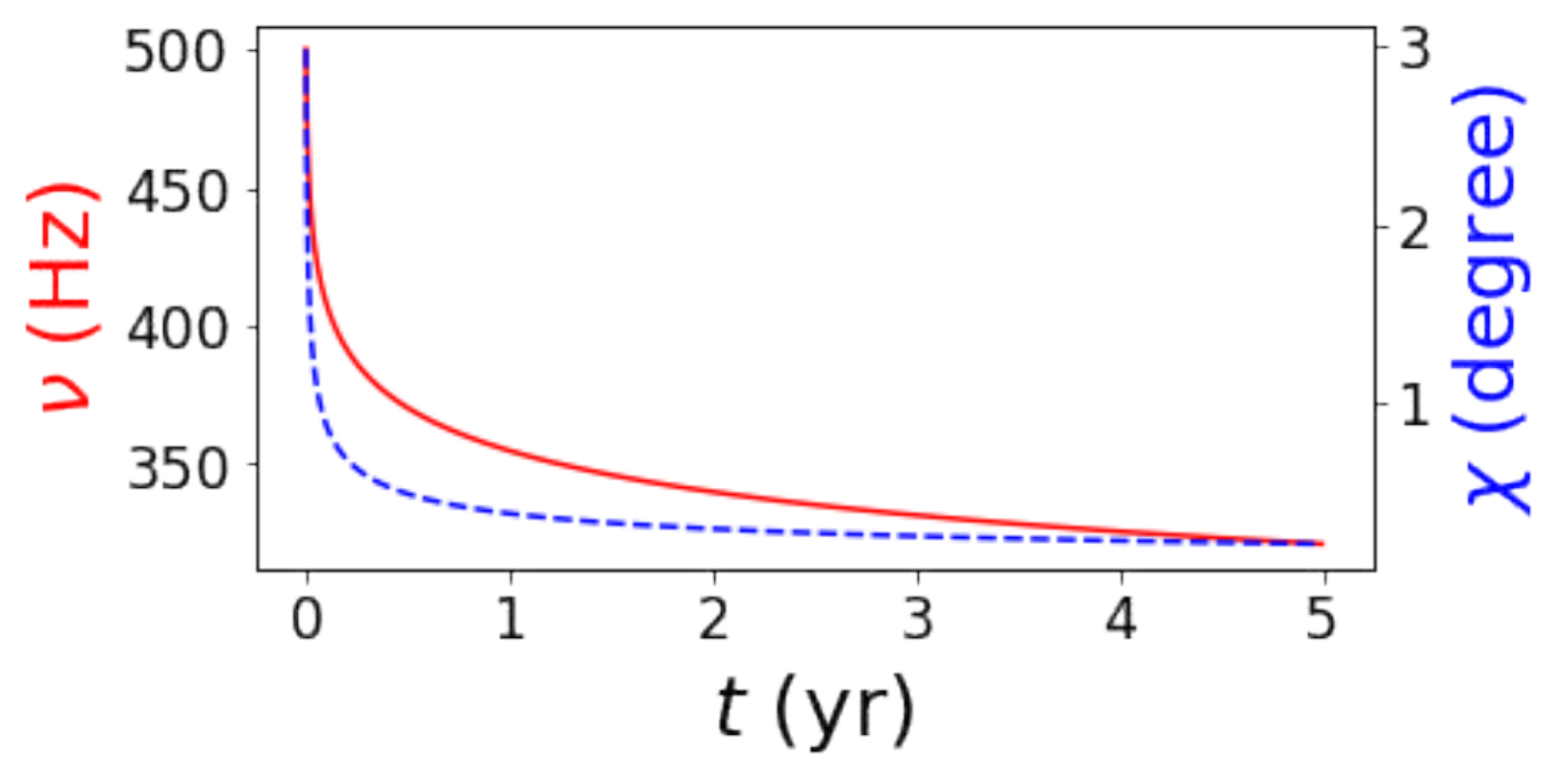}
\includegraphics[width=\columnwidth]{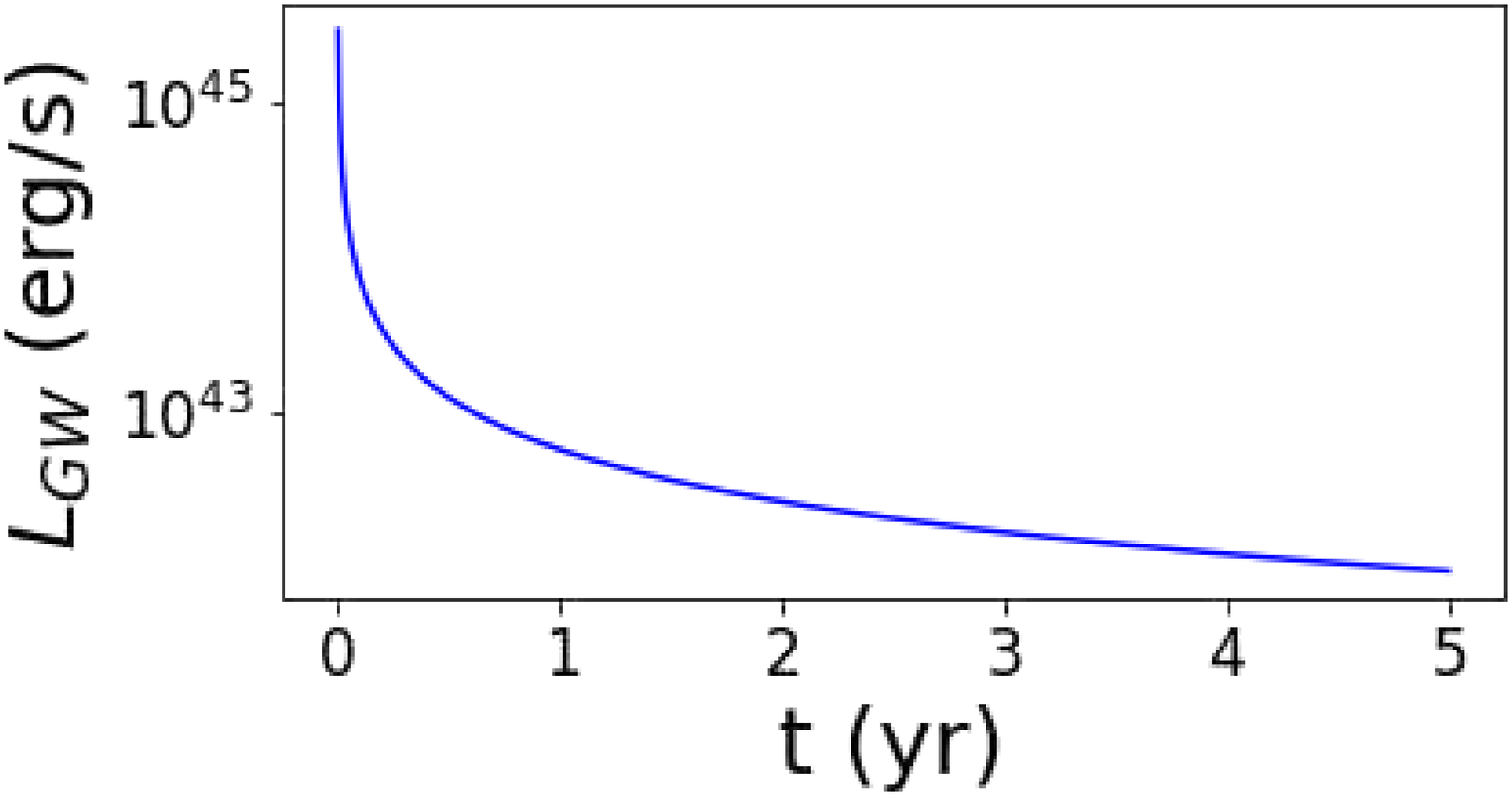}
\caption{For toroidally dominated NSs, variations of $\nu$, $\chi$ and $L_{GW}$ as 
	functions of time for $B_{max}=1.4\times 10^{17}$ G, initial $\nu=500$ Hz, $\chi=3^\circ$, as given in Table \ref{tab:torwxdecay}. Red (solid) and blue (dashed) lines in the upper panel show the variations of $\nu$ and $\chi$, respectively.}
\label{fig:torwxlgw}
\end{center}
\end{figure}

In the above results, we have simply assumed the NS to be poloidally dominated so that we can effectively use the formula for $L_D$. Actually, stable NSs consist of a suitable mixture of toroidal and poloidal components. As {\it XNS} cannot handle such configuration, now we consider a few cases of NSs for $\Gamma=1.95$ and $\rho_c = 10^{15}$ g cm\textsuperscript{-3} for purely toroidal magnetic field and drop the contribution from the term $L_D$, assuming that even if the NS possesses any dipole contribution, its effect is much smaller. As in this configuration $L_{GW} \gg L_D$, such a magnetized massive NS can radiate for a long time. Table \ref{tab:torwxdecay} shows the timescales, $T^\prime_\Omega$, and $T^\prime_\chi$, for various NSs to be active in radiating with a toroidally dominated magnetic field. 
However, detecting such NSs by the GW detector and the duration of detection depends on the (relative) strengths of the toroidal and poloidal field components. If the stable mixed field configuration is indeed toroidally dominated as suggested by, e.g., \citet{W2014}, {that configuration remains stable even after a long time, which satisfies the stability criteria given by } \citet{BR2009}. Such NSs will be detectable for a longer duration by GW detectors, as implied by Fig. \ref{fig:torwxlgw} when $L_{GW}$ remains high for a longer duration.
The dimensionless GW amplitude for NSs is given in Table \ref{tab:torwxdecay} at $t=0$ 
which, however, will decay with time due to evolutions of $\Omega$ and $\chi$.

\begin{figure}
\begin{center}
\includegraphics[width=\columnwidth]{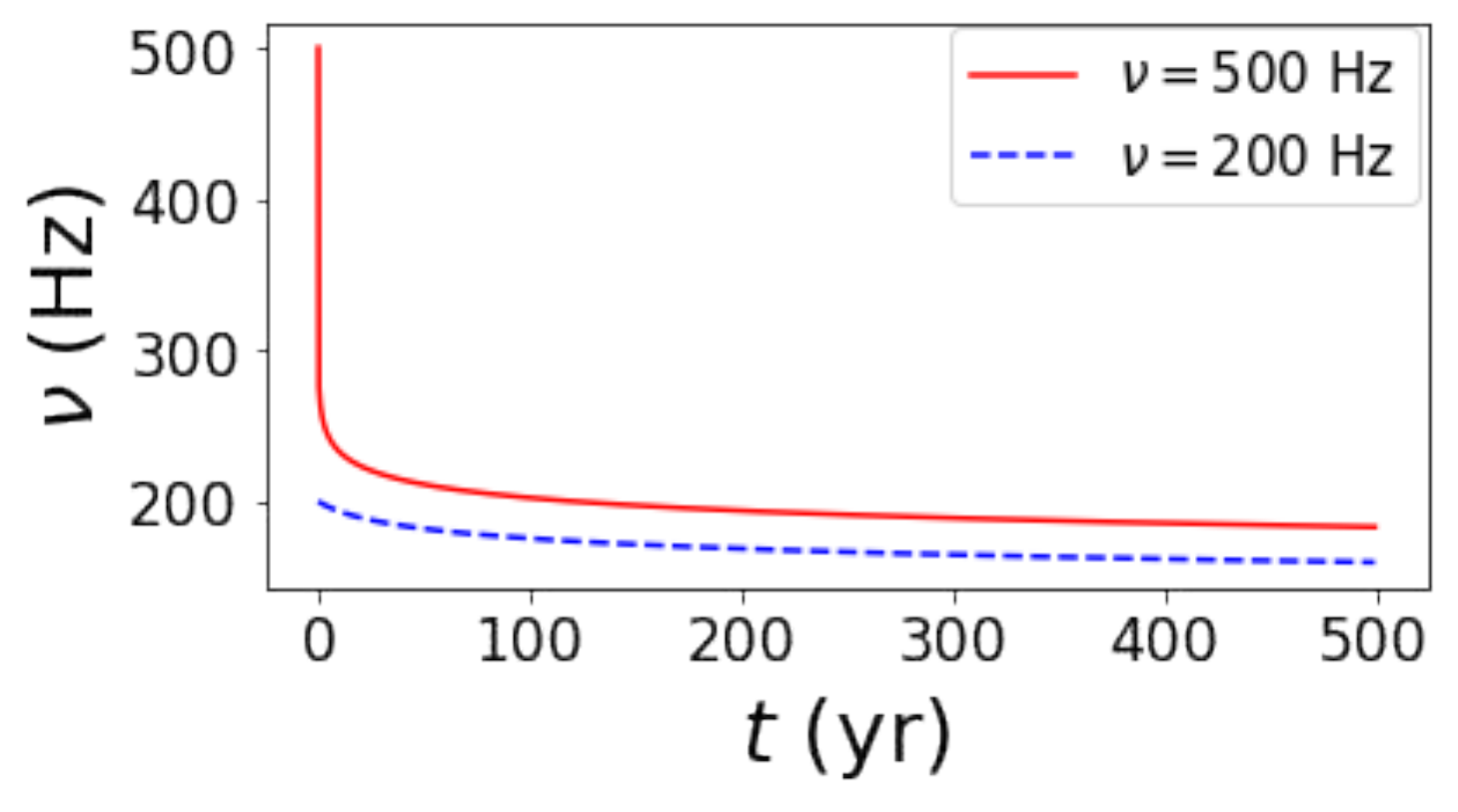}
\includegraphics[width=\columnwidth]{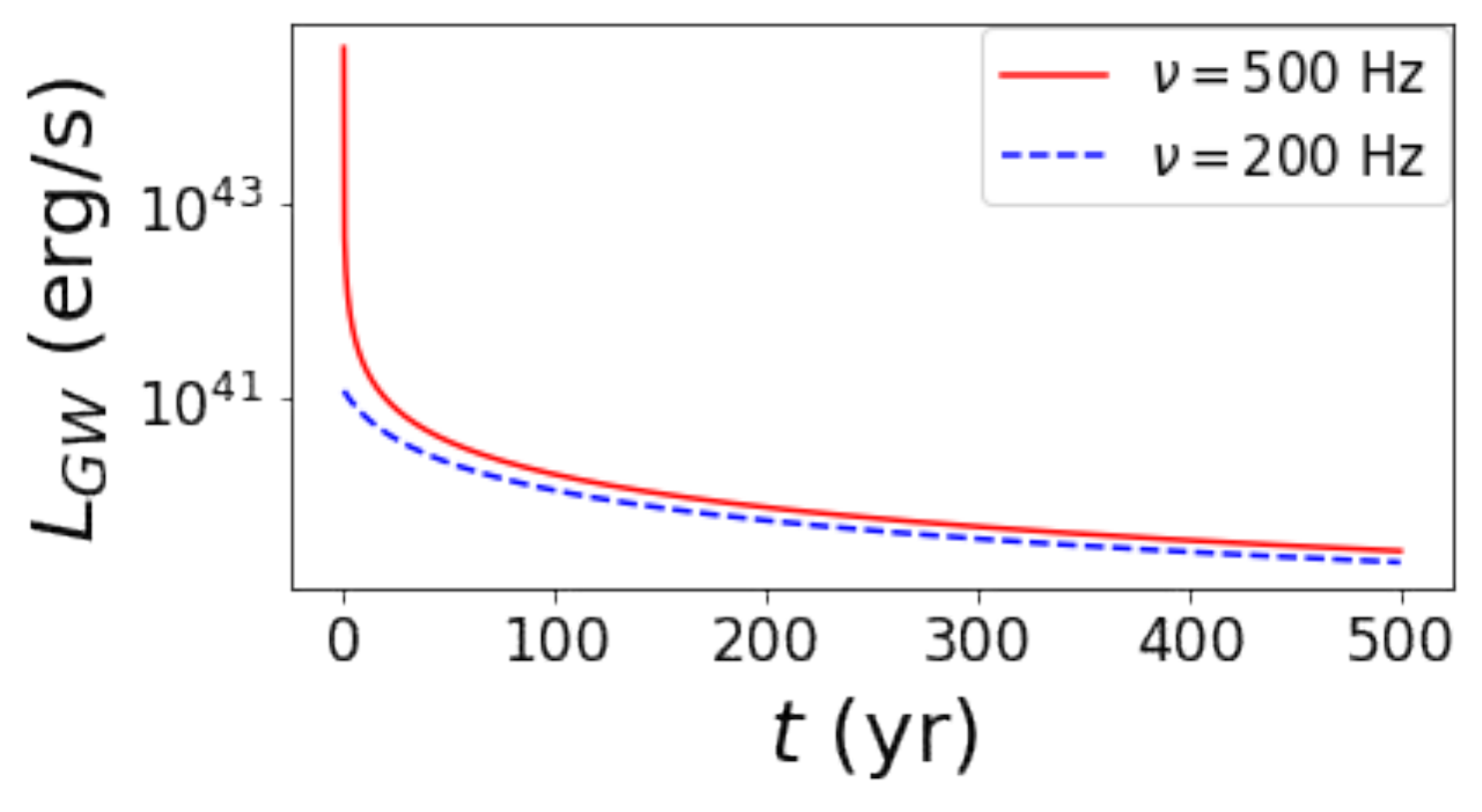}
\caption{Variations of $\nu$ and $L_{GW}$ as functions of time, for different rotation, keeping magnetic field same ($B_{max}=9\times 10^{16}$G).}
\label{fig:torvarnu}
\end{center}
\end{figure}

Also, we show in Table \ref{tab:torwxdecay} that the timescale for NSs behaving as a pulsar
increases with smaller $\Omega$ (thus smaller $\nu$). Indeed Fig. \ref{fig:torvarnu} shows that the change in
$\nu$ decreases with decreasing initial $\nu$, but $T^\prime_\Omega$ and $T^\prime_\chi$
increase for smaller initial $\nu$.

\begin{figure}
\begin{center}
\includegraphics[width=\columnwidth]{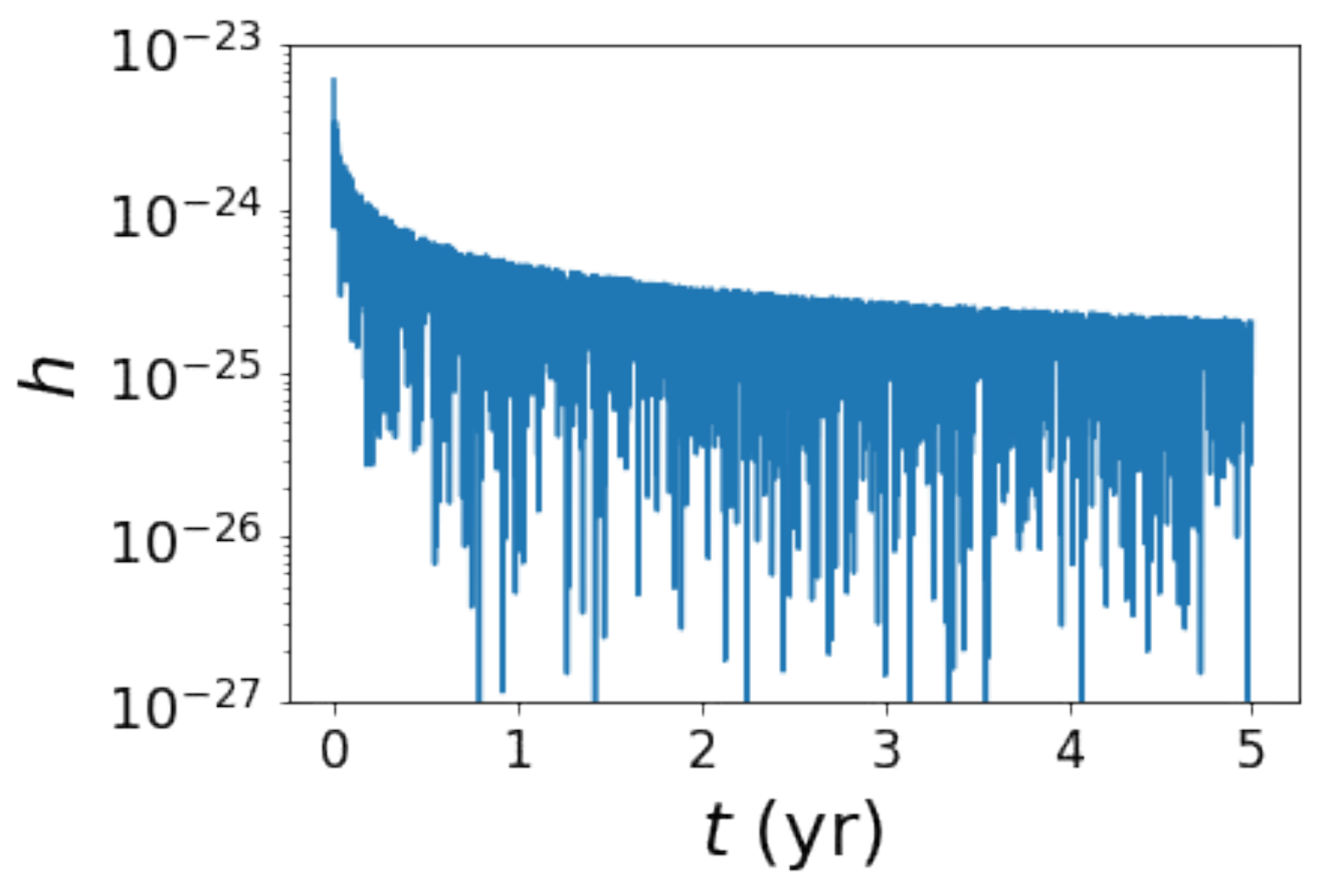}
\caption{Dimensionless GW amplitude as a function of time due to decays of $\Omega$ and $\chi$ (thus $h$) for $B_{max}=1.4\times 10^{17}$ G, initial $\nu=500$ Hz, $\chi=3^\circ$, as given in Table \ref{tab:torwxgwdecay}.}
\label{fig:torwxgwdecay}
\end{center}
\end{figure}

\begin{figure}
\begin{center}
\includegraphics[width=\columnwidth]{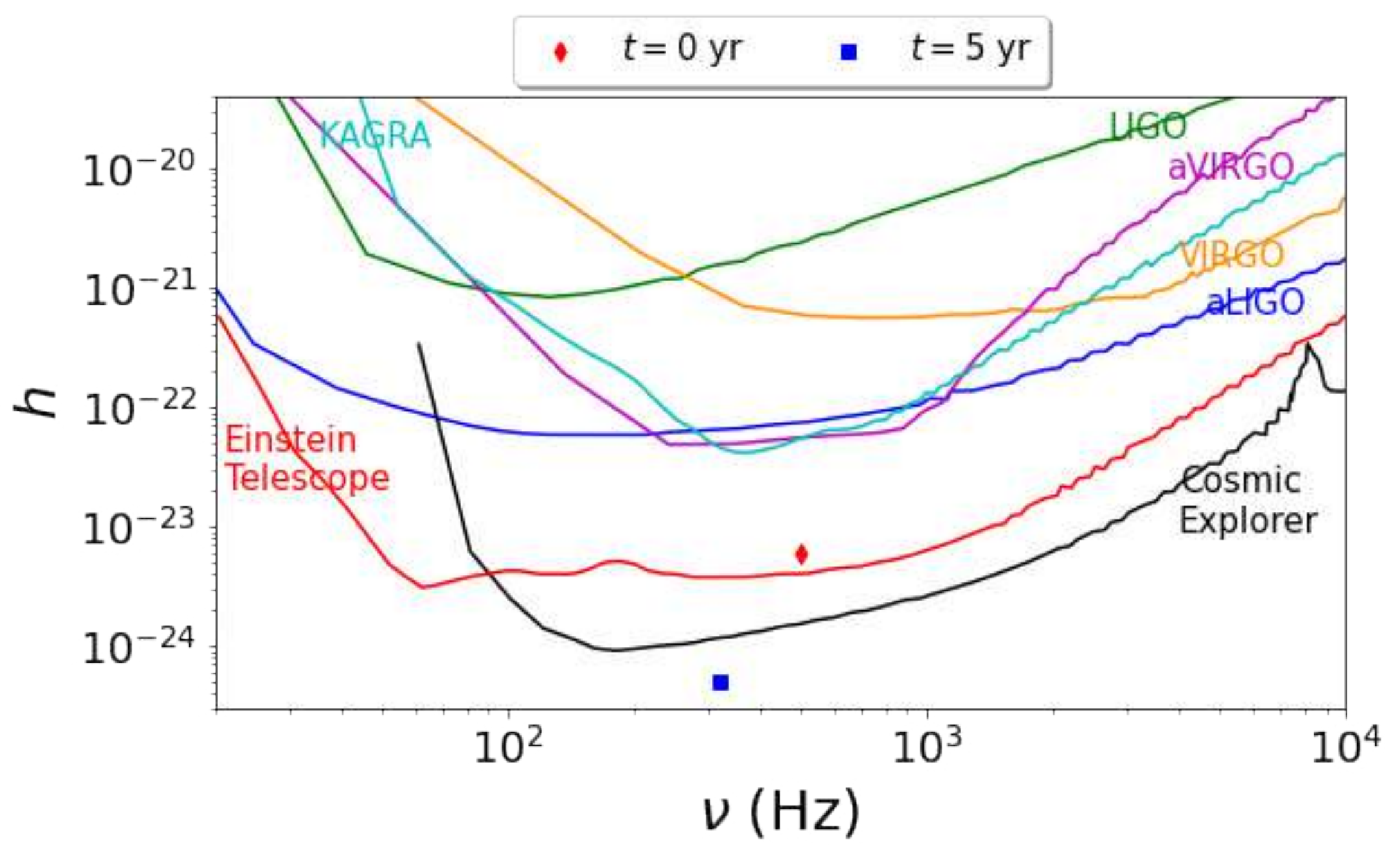}
\caption{Dimensionless GW amplitude for NSs before and after $\nu$ and $\chi$ decay along with the sensitivity curve of various detectors for $B_{max}=1.4\times 10^{17}$ G, initial $\nu=500$ Hz, $\chi=3^\circ$, as given in Table \ref{tab:torwxgwdecay}.}
\label{fig:torwxgwdecaysen}
\end{center}
\end{figure}

Dimensionless GW amplitude, which is a function of $\chi$ and $\Omega$ 
(equations \ref{eq:gwstrain} and \ref{eq:gwamp}), decays with time as well, as shown in 
Table \ref{tab:torwxgwdecay} and Fig. \ref{fig:torwxgwdecay}. The dimensionless GW amplitude for NSs at their birth and after some time, along with sensitivity curves of 
various detectors have been shown in 
Fig. \ref{fig:torwxgwdecaysen}.
as discussed in Sections \ref{sec:Bdecaytor} and \ref{sec:Bdecaypol}. Even though the magnetic field remains appreciable, due to decays of $\Omega$ and $\chi$, the GW amplitude turns out to be much smaller beyond certain time. Also, we can see by comparing timescales given in Tables \ref{tab:Bdecaytor} and 
\ref{tab:Bdecaypol} with those in Tables 
\ref{tab:polwxdecay}, \ref{tab:torwxdecay} and \ref{tab:torwxgwdecay} that the timescale for magnetic field decay is very much larger than the decay timescales of $\Omega$ and/or $\chi$. Thus, we consider the magnetic field to be constant during the decays of $\Omega$ and $\chi$.
The only exception is the poloidal case with the initial poloidal field
$10^{12}$ G given in Table \ref{tab:polwxdecay} such that $L_{GW}>>L_D$, 
when the timescale to saturate $\Omega$ is comparable to the field decay 
timescale. However, $\chi$ becomes zero
in two orders of magnitude shorter time, and once it is zero, there
is no GW emission (whose amplitude is very small in this case). 
Therefore, till the NS remained pulsar, the field was practically unchanged.
Hence, for the all practical purposes, the choice of constant magnetic field 
in the evolutions of $\chi$ and $\Omega$ for the cases presented in the 
paper is fine.

\begin{table*}
\caption{Toroidal magnetic field with $\rho_c=10^{15}$ g ${\text{cm}}^{-3}$. $I_{xx}$ and $I_{zz}$ are in the units $4.5\times10^{43}$ g cm\textsuperscript{2}.}
\label{tab:torwxdecay}
\begin{center}
\small
\begin{tabular}{c c c c c c c c c c c c c c} 
\hline\hline
$M$&$R_E$&$B_{max}$& $\nu$& ME/GE & KE/GE & $I_{xx}$ & $I_{zz}$ & $L_{GW}$ &$h_0~(d=10~kpc)$& $T^\prime_\Omega$& $T^\prime_\chi$\\
$(M_{\odot})$ & (km)&(G) &(Hz)& & & & &(erg/s) & at $t=0$ &(yr) &(yr)\\
\hline
1.963 & 12.65 & $1.4\times 10^{17}$ & 500 & $4.9\times 10^{-3}$ & $2\times 10^{-2}$ & 11.53 & 12.22 & $2.9\times10^{45}$ & $5.6\times 10^{-22}$ & 0.016 & 0.006 \\

1.963 & 12.48 & $9\times 10^{16}$ & 500 & $2\times 10^{-3}$ & $2\times 10^{-2}$ & 11.411 & 11.156 & $3.9\times10^{44}$ & $2.2\times 10^{-22}$ & 0.1 & 0.04 \\

1.909 & 11.99 & $9\times 10^{16}$ & 200 & $2\times 10^{-3}$ & $3.1\times 10^{-3}$ & 11.47 & 11.54 & $1.2\times10^{41}$ & $3.5\times 10^{-23}$ & 56.5 & 20.0 \\
\hline
\end{tabular}
\end{center}
\end{table*}

\begin{table*}
\caption{Change of GW strain ($h$) for toroidally dominated NS with $\rho_c=10^{15}$ g ${\text{cm}}^{-3}$, $\Gamma=1.95$, due to $\Omega$ and $\chi$ decay.}
\small
\begin{tabular}{c c c c c c c c c c c c} 
\hline\hline
t(yr) & $M$$(M_{\odot})$&$R_E$(km) & $B_{max}$ (G) & $\nu$ (Hz) & ME/GE & KE/GE & $|\epsilon|$ & $h~(d=10~kpc)$\\
\hline
0 & 1.963 & 12.65 & $1.4\times 10^{17}$ & 500 & $4.9\times 10^{-3}$ & $2\times 10^{-2}$ & 0.01 & $6\times 10^{-24}$\\
5 & 1.924 & 12.32 & $1.4\times 10^{17}$ & 320 & $4.8\times 10^{-3}$ & $8\times 10^{-3}$ & 0.01 & $5\times 10^{-25}$ \\
\hline
\end{tabular}
\label{tab:torwxgwdecay}
\end{table*}

\section{Viscous and thermal effects on obliquity angle}
\label{sec:vist}
 
In all the above discussions, we have studied the evolution of obliquity angle due to electromagnetic and GW radiations, both of which extract the angular momentum from the star. As a result, the obliquity angle decreases, leading to the alignment of the magnetic and rotating axes of the star. However, in the early days after the formation of NSs before transitioning to superfluidity, the viscous effect may try to increase the obliquity angle by redistributing angular momentum.

This may turn the NS into an orthogonal rotator if the NS has a prolate deformation \citep{CC2002} due to a toroidally dominated magnetic field. 
 
The early evolution of the obliquity angle depends on the relative strength between dipole/quadrupole radiations and viscous damping, which in turn depends on the respective strengths and geometries of the magnetic fields. 

If the viscous damping is dominating for a particular model, the star will be able to radiate GW for a longer time, which will help them to be detected, as a larger $\chi$ leads to higher GW amplitude \citep{CC2002,Dall2009}.

The viscous effect arises due to bulk viscosity, as the shear viscosity effect is negligible in NS in order to affect $\chi$.
The bulk viscosity effect arises due to departure from chemical equilibrium when the matter is compressed and expanded due to the perturbation by Urca processes \citep{landau1987,lind2002}, the corresponding timescale is given by \citep{LJ2018}

\begin{eqnarray}
\nonumber
&& T_{bulk}= 7.9 \text{s} \left(\frac{M}{1.4M_\odot}\right)^{5/3} \left(\frac{R}{10km}\right)^{-1} \left(\frac{T}{10^{10}K}\right)^{6}\\ &&\left(\frac{\nu}{khz}\right)^{-4}
\left(\frac{B}{10^{15}G}\right)^{-2} \left(\frac{\sin^2\chi}{g(\chi)}\right),
\end{eqnarray}
for $\omega\tau<<1$, and
\begin{eqnarray}
\nonumber
&& T_{bulk}= 0.19 \text{s} \left(\frac{M}{1.4M_\odot}\right)^{-1} \left(\frac{R}{10km}\right)^2 \left(\frac{T}{10^{10}K}\right)^{-6}\\ &&\left(\frac{\nu}{khz}\right)^{-2}
\left(\frac{B}{10^{15}G}\right)^2 \left(\frac{\sin^2\chi}{g(\chi)}\right),
\end{eqnarray}
for $\omega\tau>>1$, where
\begin{eqnarray}
\nonumber
&&\omega\tau=0.14\times \cos\chi \left(\frac{\nu}{khz}\right) \left(\frac{B}{10^{15}G}\right)^2 \left(\frac{R}{10km}\right)^2\\
&&\left(\frac{T}{10^{10}K}\right)^{-6} \left(\frac{M}{1.4M_\odot}\right)^{-4/3}.
\end{eqnarray}
and $g(\chi) \sim \sin^2\chi$ for $\chi<<1$ and $g(\chi)\sim \cos^2\chi$ for $\pi/2-\chi<<1$. Note importantly that the timescale depends on the temperature.

he early evolution due to viscosity, we will focus on bulk viscosity. 

In an early evolution of NS, the cooling due to the Urca process \citep{owen1998,page2006} is also important to include as the viscosity coefficients are temperature dependent and the star cools down from the initial temperature (at the end of the proto-NS phase) of around $10^{11}$ to $10^{9}$ K in days. The modified-Urca cooling, given by \citet{page2006} assuming the matter is too hot to become superconductor and/or superfluid, leads to the
temperature profile as
\begin{equation}
T(t)=\left(\frac{6N^s}{C}t+\frac{1}{T_0^6}\right)^{-1/6},
\end{equation}
where $T_0$ is the temperature at time $t = 0$, and the constants $N_s = 10^{-32}$ s\textsuperscript{-1} K\textsuperscript{-8},
$C = 10^{30}$ erg K\textsuperscript{-2}
and $T_0 = 10^{11}$ K.
The characteristic timescale for cooling is given by \citet{page2006} as,
\begin{equation}
T_{cooling}=16 \left(\frac{C}{10^{30}}\right)  \left(\frac{N_s}{10^{-32}}\right)^{-1} \left(\frac{T}{10^{10}}\right)^{-6} \text{s} ,
\end{equation}

Therefore, the most general evolution equation for $\chi$, including the effects of bulk viscosity, could be \citep{landau1987,lind2002,LJ2017} given by

\begin{eqnarray}
\nonumber
&&I_{z^{\prime}z^{\prime}}\frac{d\chi}{dt}=-\frac{12G}{5c^5} (I_{zz}-I_{xx})^2 \Omega^4 \sin^3 \chi \cos \chi\\
\nonumber
&&-\frac{B_p^2 R_p^6 \Omega^2}{2c^3} \times \sin \chi \cos \chi F(x_0)
+\zeta {\epsilon_{\Omega}}^2 \epsilon R^3 \frac{g(\chi) }{I_{zz} \sin\chi \cos\chi},
\label{dchidtvis}
\end{eqnarray}

\begin{eqnarray}
\nonumber
\epsilon_\Omega=0.21 \left(\frac{R}{10km}\right) \left(\frac{\nu}{khz}\right)^2 \left(\frac{M}{1.4M_\odot}\right)^{-1},
\nonumber
\end{eqnarray}

\begin{eqnarray}
\nonumber
&&\zeta=4.2\times10^{33}\text{g cm\textsuperscript{-1} s\textsuperscript{-1}} \left(\frac{T}{10^{10}K}\right)^{-6}\\
&&\left(\frac{M}{1.4M_\odot}\right)^{10/3} \left(\frac{R}{10^6}\right)^{-10}, 
\end{eqnarray}
for $\omega\tau<<1$, and

\begin{eqnarray}
\nonumber
&&\zeta=1.7\times10^{35}\text{g cm\textsuperscript{-1} s\textsuperscript{-1}} \left(\frac{T}{10^{10}K}\right)^6 \left(\frac{M}{1.4M_\odot}\right)^6\\
&& \left(\frac{B}{10^{15}G}\right)^{-4} \left(\frac{\nu}{khz}\right)^{-2} \left(\frac{R}{10km}\right)^{-14} \left(\frac{1}{\cos^2\chi}\right), 
\end{eqnarray}
for $\omega\tau>>1$. The above equation, along with the evolution of $\Omega$ (which remains unchanged as before), needs to be solved for the full time evolution of the obliquity angle.

Here bulk viscosity is assumed for a non-superfluid NS made only of neutrons, protons, and electrons \citep{LJ2018}. However, below a certain temperature, the hyperons may appear in the core when it becomes superfluid, which significantly increases the bulk viscosity \citep{jones1976,lind2002}, hence the calculation is most conservative.

As $\zeta$ and the corresponding effect in the evolution of $\chi$ steeply depend on $T$, due to the cooling of proto-NS, effects of $\zeta$ suppress very fast. Therefore, practically NSs under consideration hardly become an orthogonal rotator and, hence, electromagnetic and GW radiations play the main role to evolve $\chi$ as discussed in the previous section.

However, if $T$ would remain constant for a long time, the obliquity angle could increase and reach $90^\circ$. 
If the NS has a pure toroidal magnetic field, e.g., $B_{max}^{tor}\sim 9\times 10^{16}$ G (and much smaller poloidal magnetic field $B_{max}^{pol}\sim 5\times10^{14}$ G), it would tend to become an orthogonal rotator in a smaller timescale (e.g., in a few hours) due to viscous effect, when $T=10^{10}$ K, compared to the time the effects due to electromagnetic and gravitational radiations would take to make it an aligned rotator (e.g., $\sim 50$ years). However, eventually, in a long run, $\chi$ would only decrease when electromagnetic and/or gravitational effects start playing a role, stopping any GW and electromagnetic radiation.

\begin{figure}
\begin{center}
\includegraphics[width=\columnwidth]{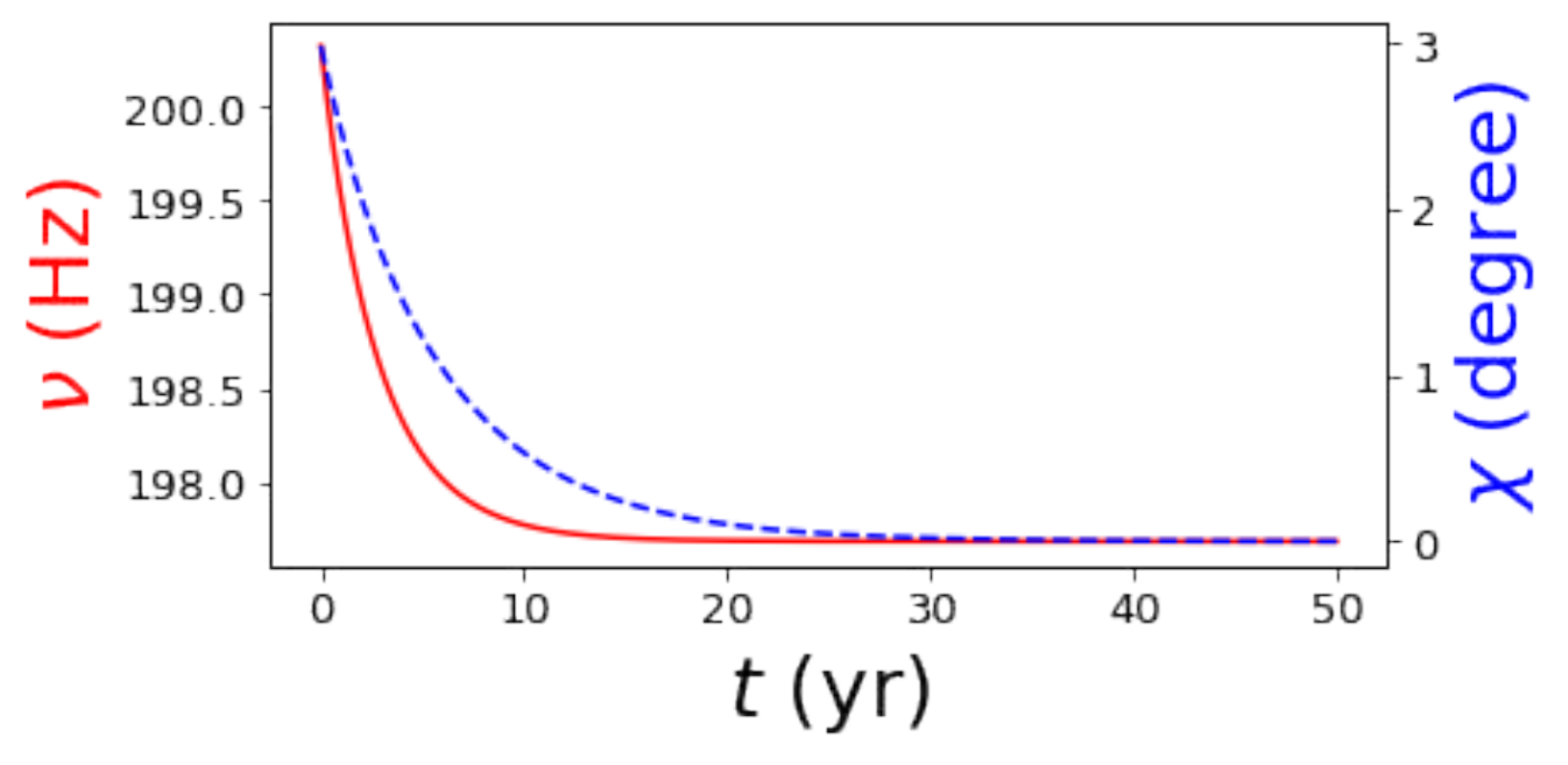}
\includegraphics[width=\columnwidth]{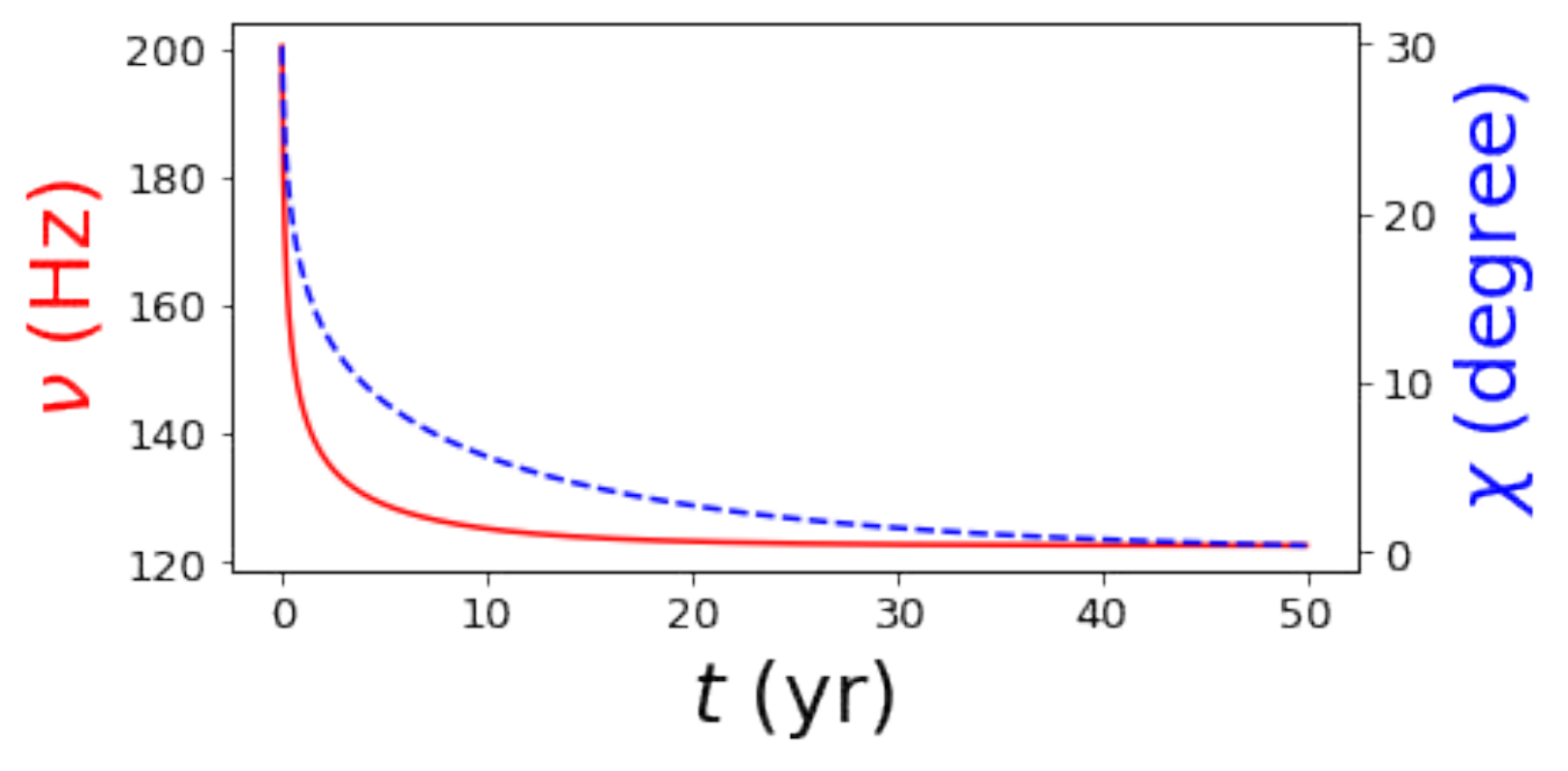}
\caption{Variations of $\nu$ and $\chi$ with time for $B_{max}^{tor}\sim 9\times 10^{16}$ G (and $B_{max}^{pol}\sim 5\times10^{14}$ G), initial $\nu=200$ Hz, and $\chi=3^\circ$ (upper panel) and $\chi=30^\circ$ (lower panel). Red (solid) and blue (dashed) lines show the variations of $\nu$ and $\chi$, respectively.}
\label{fig:timescale30}
\end{center}
\end{figure}

We also explore, for completeness, the $\chi$ evolution due to electromagnetic and/or gravitational radiations only with its initial value $30^\circ$. Fig \ref{fig:timescale30} shows that the timescale changes between the initial $\chi=3^\circ$ and $30^\circ$ only by a few factors. \\

\section{Sensitivity of continuous gravitational wave detectors}
\label{sec:snr}

We study the plausibility of instantaneous GW detection from isolated NSs by some existing and upcoming detectors. At present, extensive effort is going on to increase the sensitivity of detectors to detect CGWs emitted from various sources \citep{SB2019}. This can be done by calculating the SNR, thereby estimating the necessary observation time for the particular GW detector to detect these objects. 
A NS behaving like a pulsar can radiate CGWs at two frequencies. {If the strength of the GW signal from the NS remains nearly unchanged during the observation time $T$, the cumulative SNR of the detector can be calculated coherently by discretizing the timescale finely, including information about phase for each grid  \citep{JA1998,BVM2010}, which takes very large computational time. However, in reality, the spin-down is so fast that $\nu$ and $\chi$ change rapidly; thus, GW strain also changes fast enough. In such a situation, the time integration is accomplished in time-stacks $T_{stack}$ such that, in each stack, $\nu$ and $\chi$ remains nearly constant. The SNR is calculated coherently for each stack and then added incoherently to obtain the cumulative SNR. The total observation time $T$ is divided into $\mathcal{N}$ time-stacks such that $T=\mathcal{N} T_{stack}$. Also, in such a technique the phase information between the different stacks gets lost, thus it is called the incoherent search (see \citealt{MAG2008}, for details of the technique). However, an incoherent search with a time-stacking method is computationally efficient compared to the coherent search \citep{BC2000,CC2005} which is used for blind search of unknown pulsars \citep{Leaci2012}.} Assuming $\nu$, $\chi$ and $h_0$ remain nearly constant over each time-stack, adding $\mathcal{N}$ such stacks, the cumulative SNRs is given by \citet{MAG2008}, as
\begin{equation}
\langle S/N\rangle =\sqrt{{\langle S/N_\Omega^2 \rangle}+{\langle S/N_{2 \Omega}^2 \rangle}},
\label{eq:snr}
\end{equation}
where 
\begin{equation}
\langle S/N_\Omega^2 \rangle = \frac{\sin^2 \zeta}{100} \frac{h_0^2 \sqrt{\mathcal{N}} T_{stack} \sin^2 2\chi}{S_n(\nu)}=\frac{\sin^2 \zeta}{100} \frac{h_0^2 T \sin^2 2\chi}{\sqrt{\mathcal{N}} S_n(\nu)}
\label{eq:snrw}
\end{equation}
and

\begin{multline}
\langle S/N_{2 \Omega}^2 \rangle = \frac{4 \sin^2 \zeta}{25} \frac{h_0^2 \sqrt{\mathcal{N}} T_{stack} \sin^4 \chi}{S_n(2\nu)}\\
=\frac{4 \sin^2 \zeta}{25} \frac{h_0^2 T \sin^4 \chi}{\sqrt{\mathcal{N}} S_n(2\nu)},
\label{eq:snr2w}
\end{multline}

where $\zeta$ is the angle between the interferometer arms and $S_n(\nu)$ is the detector's power spectral density (PSD) at the frequency $\nu$ with $\Omega=2\pi \nu$. The data for PSD of various detectors are extracted from \citet{MOORE2014} and \citet{H2020}. For ground-based interferometers such as LIGO, VIRGO, KAGRA, Cosmic Explorer, etc. $\zeta=90^\circ$ and for space-based interferometers such as Einstein Telescope $\zeta=60^\circ$. Note that the average is over all possible angles, including $i$, which determines the object's orientation with respect to the celestial sphere reference frame.

{Note that, in such a stacking technique, the SNR reduces by a factor $\mathcal{N}^{1/4}$ compared to the continuous integration in a fully coherent search. Here we use the stacking method with $T_{stack} \approx 3$ hr. Also, in such a long time, the antenna pattern may change with time, and there might be movement of the antenna itself, which may lead to a change in SNR, which will make the computation even more challenging \citep{MAG2008}. Hence, while SNR $\sim 11$ is the appropriate threshold for the short-duration binary inspiral, due to the larger effective number of templates for the continuous search, it is computationally more challenging. This suggests to increase the threshold value of SNR \citep{DP2021} to $>11.4$ \citep{AB2003,MAG2020,CBT2021} for CGW; thus, we set the threshold value for SNR to be 12 for more than $95\%$ detection efficiency \citep{R2012,AB2016}.}

\subsection{Possible detection of massive poloidally dominated neutron star pulsars}

\begin{figure}
\begin{center}
\includegraphics[width=\columnwidth]{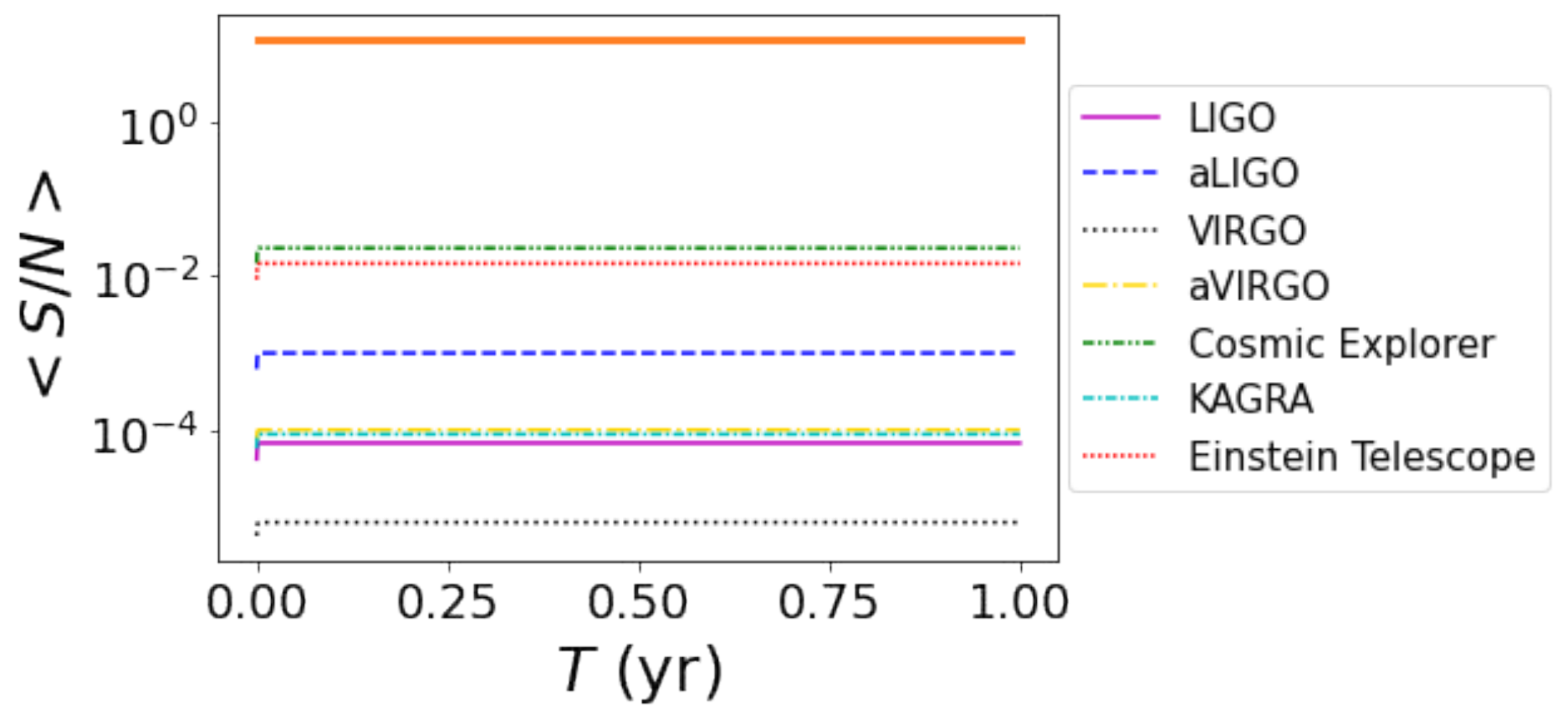}
\includegraphics[width=\columnwidth]{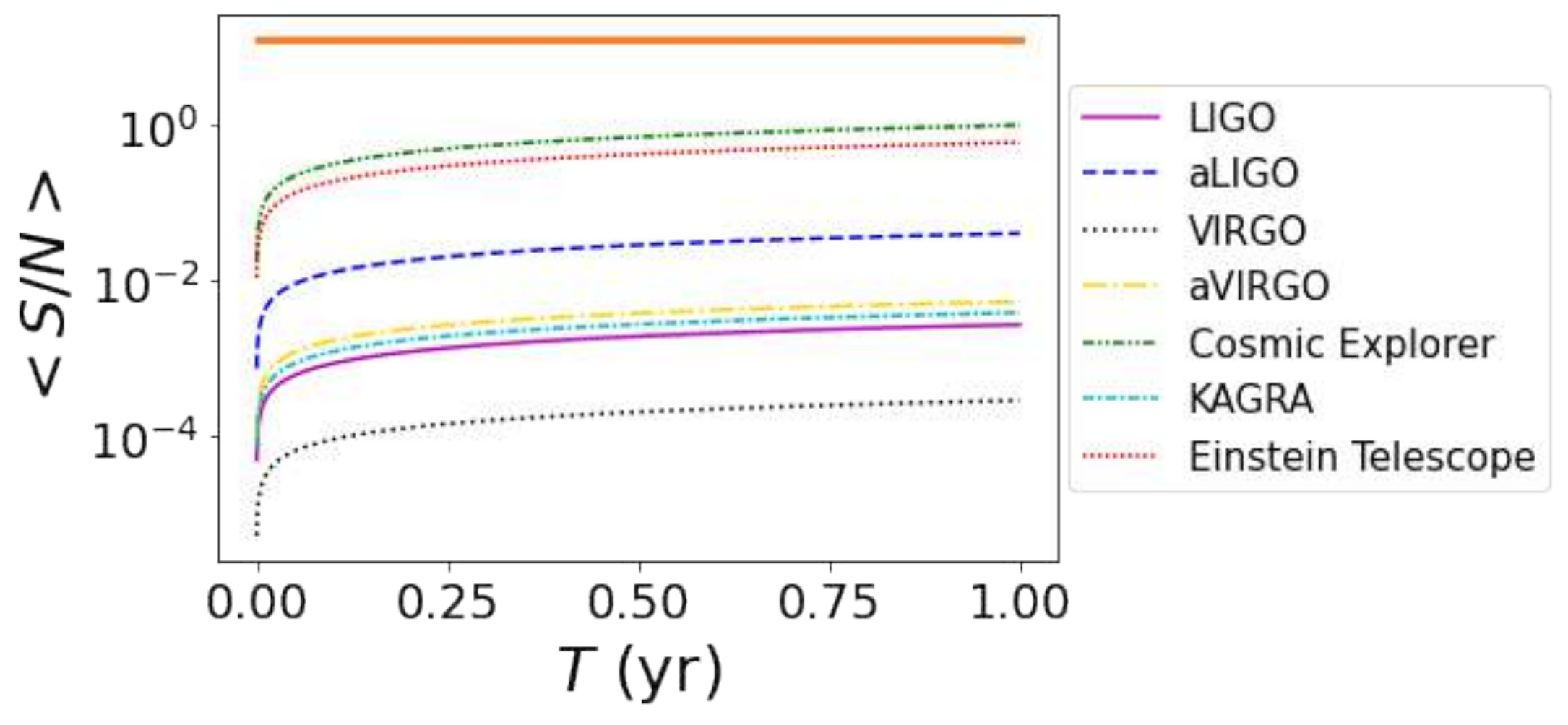}
\caption{SNR for various detectors as a function of integration time for a poloidal magnetic field dominated NS with initial $\nu=100$ Hz, $\chi=3^\circ$, $B_P=10^{15}$ G ($L_D>l_{GW}$; top panel) and $B_P=10^{12}$ G ($L_{GW}>L_D$; bottom panel), for two cases from Table \ref{tab:polwxdecay}. The orange line corresponds to $<S/N>=12$.}
\label{fig:snrpolB}
\end{center}
\end{figure}

We solve equations (\ref{eq:dwdt}) and (\ref{eq:dxdt}) simultaneously to obtain $\Omega(t)$ and $\chi(t)$ assuming poloidal field dominated NSs and then calculate the cumulative SNR for various detectors \citep{KMMB2020,KMTBM2021}. Fig. \ref{fig:snrpolB} shows the SNR as a function of time for a poloidal field dominated NSs with different field strengths at the pole, i.e. $10^{15}$ G and $10^{12}$ G. $B_P$ is larger in the first case; thus, $\Omega$ and $\chi$ decrease rapidly with time due to the large $L_D$. In the stacking method, the power of the GW signal for each stack is added up; thus, SNR increases for about one month and eventually saturates, as seen in Fig. \ref{fig:snrpolB}. This happens because when $\Omega$ and $\chi$ decrease significantly, the strength of GW amplitude also decreases; thus, the power for later stacks decreases. Hence, adding more stacks with comparatively less power does not efficiently change the cumulative SNR. However, when $B_P$ is smaller, $L_D$ is lower, and the SNR always increases with time for $1$ yr because both $\Omega$ and $\chi$ remain nearly constant over the integration time, as seen in Fig. \ref{fig:snrpolB}. It is, however found that none of the detectors will be able to detect such NSs even after 1 yr of integration time. This is because the magnetic field $\approx 10^{14}$ is tiny for NS to produce sufficient deformation for GW radiation. The rotation rate $\nu=100$ Hz is also quite small, which affects GW amplitude ($h_0\propto\epsilon\Omega^2$). Now, if we increase $B_P$ which is around $1.2\times10^{17}$ G  (Table \ref{tab:polwxdecay}), although $h_0$ increases enough to be detectable instantaneously by most of the detectors, $\Omega$ and $\chi$ decay in a couple of seconds. Hence, it should be very rare to detect such NSs, and there is no point in calculating cumulative SNR, which will always be saturated. If we increase $\nu$ for $B_P=1.2\times 10^{14}$ G, although $h_0$ increases, which might help increase SNR, $\Omega$ and $\chi$ decay faster, hence SNR saturates faster; finally, SNR of any detector will not be able to increase up to 12.\\

\subsection{Possible detection of massive toroidally dominated neutron star pulsars}

\begin{figure}
\begin{center}
\includegraphics[width=\columnwidth]{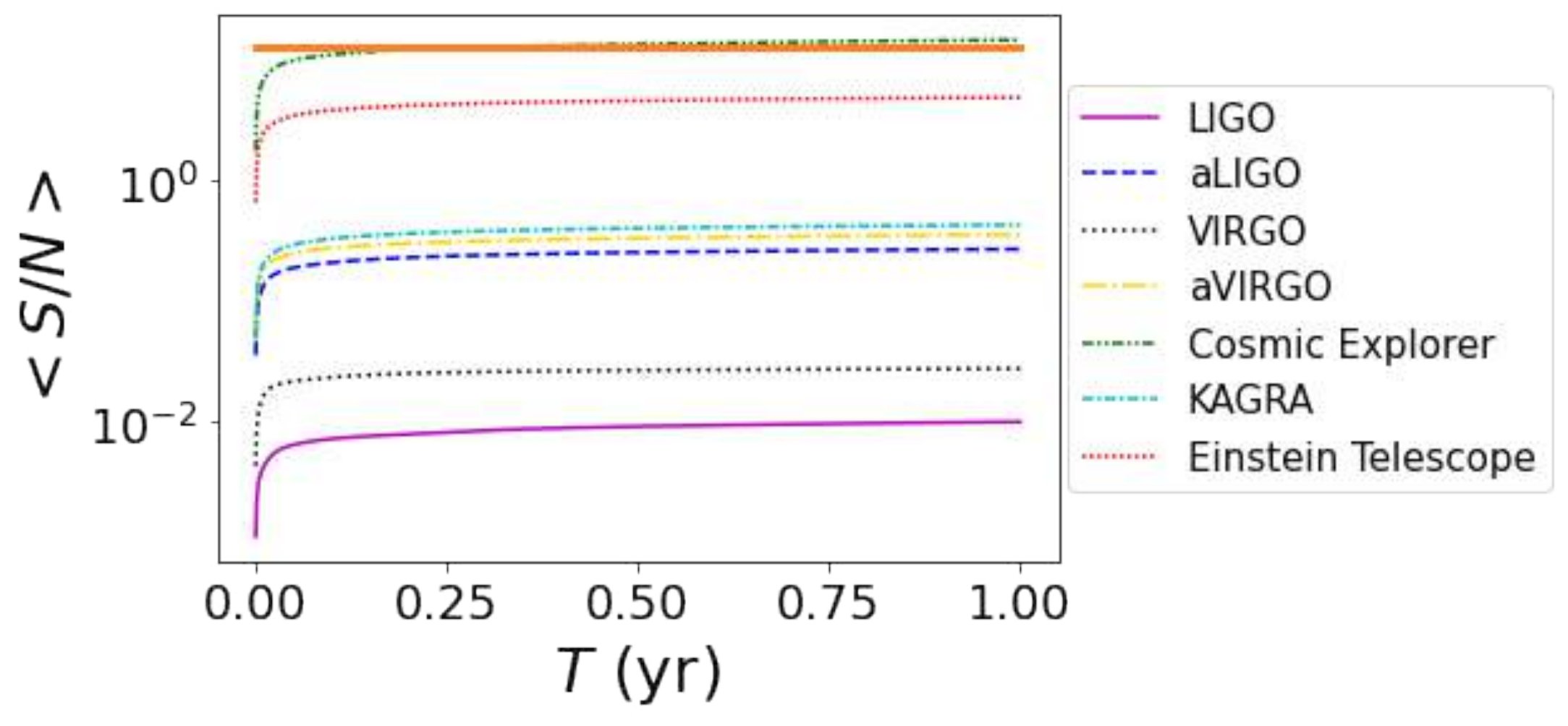}
\includegraphics[width=\columnwidth]{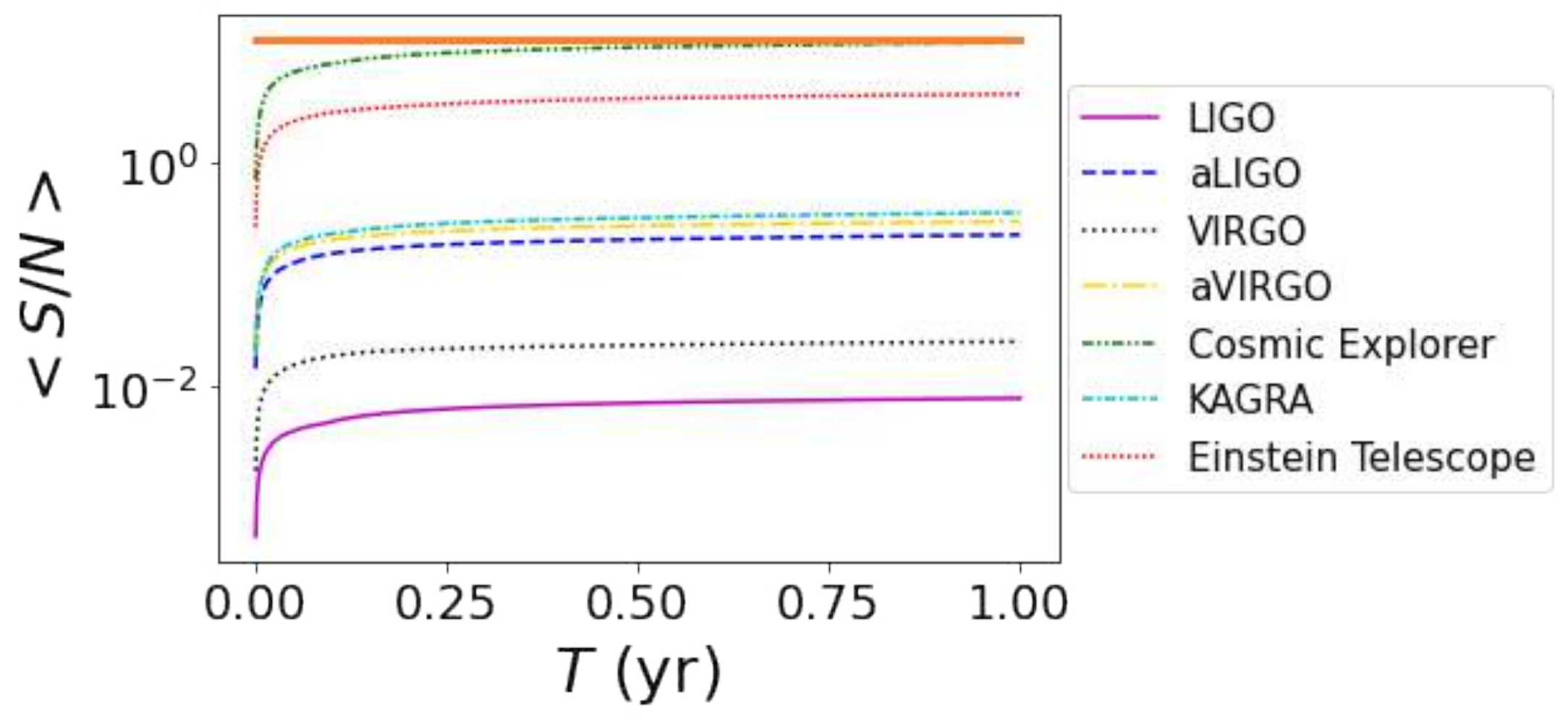}
\caption{SNR for various detectors as a function of integration time for a toroidal magnetic field dominated NS with initial $\chi=3^\circ$, initial $\nu=500$ Hz, $B_{max}=1.4\times 10^{17}$ G (top panel) and $B_{max}=9\times 10^{16}$ G (bottom panel), for two cases from Table \ref{tab:torwxdecay}. The orange line corresponds to $<S/N>=12$.}
\label{fig:snrtorB}
\end{center}
\end{figure}

\begin{figure}
\begin{center}
\includegraphics[width=\columnwidth]{0522snr.pdf}
\includegraphics[width=\columnwidth]{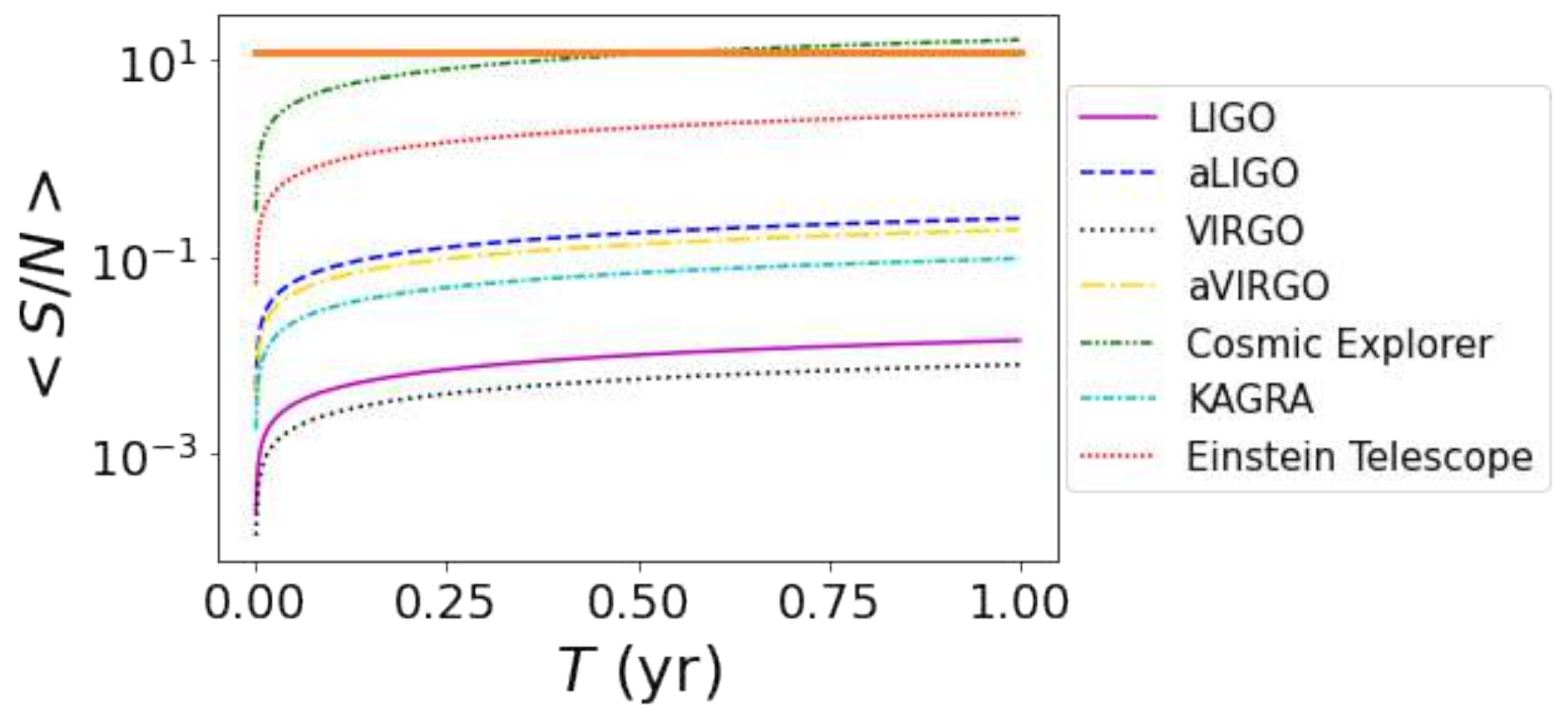}
\caption{SNR for various detectors as a function of integration time for a toroidal magnetic field dominated NS with initial $\chi=3^\circ$, $B_{max}=9\times 10^{16}$ G, initial $\nu=500$ Hz (top panel) and $\nu=200$ Hz (bottom panel), for two cases from Table \ref{tab:torwxdecay}. The orange line corresponds to $<S/N>=12$.}
\label{fig:snrtorrot}
\end{center}
\end{figure}

Fig. \ref{fig:snrtorB} shows the SNR as a function of time for toroidal field-dominated NSs with different field strengths. As \textit{XNS} cannot handle suitable toroidally dominated mixed field configuration with rotation, we assume toroidal dominated NSs with a poloidal surface field which is negligible with respect to the maximum toroidal field $B_{max}$. Such a poloidal field cannot change the shape and size of the NS as effectively as the toroidal field. Therefore, we run it for purely toroidal magnetic fields to obtain the shape and size of the NS. As $L_D$ is small, $\Omega$ and $\chi$ hardly change within a 1 yr period. Fig. \ref{fig:snrtorB} shows the SNR for NSs with $B_{max}=1.4\times 10^{17}$ G and $9\times 10^{16}$ G. {For the first case, Cosmic Explorer will be able to detect it in some months of integration. For the second case, none of the detectors will be able to detect such NSs even after one year of integration because, When the field strength decreases, the SNR decreases, as shown in Fig. \ref{fig:snrtorB}. If we increase the rotation rate, $\Omega$ and $\chi$ decay faster; thus, SNR saturates faster. For initial $\nu=200$ Hz, the SNR increases over 1 yr because the $\Omega$ and $\chi$ decrease more slowly than that of $\nu=500$ Hz, which is shown in Fig. \ref{fig:snrtorrot}, and thus, the (relatively) slowly rotating NS might be detectable by Cosmic Explorer while the (relatively) fast rotating NS will not be detectable by any of the detectors.}

\section{Magnetic Braking: Mass loss of differentially rotating neutron stars}
\label{sec:alfven}

Differentially rotating NSs can support significantly more mass than their uniformly rotating counterparts. The remnant of a binary NS merger or core collapse in a supernova may produce a hypermassive differentially rotating NS \citep{SU2000,ZM1997,RMR1998} where the remnant's core rotates considerably faster than its equator. {However, magnetic braking forces the fluid to rotate as a rigid body while the vertical field is twisted, creating a powerful toroidal field. Then, the built-up magnetic stress back-reacts to drive differential rotation in the opposite direction, unwinding the magnetic field and then winding it up again in the opposite sense. This process will create Alfv\'en waves; the toroidal field component oscillates back and forth in a standing Alfv\'en wave pattern in the Alfv\'en timescale. The angular velocity profile oscillates around a state of uniform rotation, with uniform rotation taking place when the toroidal magnetic field is at maximum magnitude. At these times, a considerable amount of the rotational energy is converted to toroidal magnetic field energy.} 

{In the presence of even small internal energy dissipation, e.g., by viscosity, the differentially rotating star will achieve a permanent state of uniform rotation by damping such dynamical oscillation. This is because the uniform rotation is the lowest energy state at a fixed angular momentum. The viscosity, along with Alfv\'en waves, carries away a significant amount of the angular momentum, and also, some amount of magnetic energy will be converted into heat. This uniform rotation configuration cannot support the excess mass anymore, and the star possibly will undergo some mass loss due to core contraction and ejection of matter in the outer envelope to form a diffused ambient disk or trigger ejection of wind \citep{SHAP2000,CSS2003,LS2003}}. This magnetic braking and angular momentum transport happen in the very first stage (a few tens of seconds) of proto-NSs, thus all radio pulsars are likely to be uniformly rotating.

{The Alfv\'en timescale depends only on the strength of the seed field. Specifically, even with a weak initial poloidal field, along with viscosity, the azimuthal magnetic field will grow to a sufficiently high value, which is adequate to brake the differential motion and drive it to uniform rotation. The Alfv\'en timescale at each radial point of the star is given by}

\begin{equation}
T_a(r)=\frac{r}{v_a(r)}=\frac{r\sqrt{4\pi \rho(r)}}{B(r)},
\end{equation}
where the density $\rho$ and field $B$ are functions of the radius, and thus Alfv\'en speed, $v_a$, is also a function of radius. We have the profiles for $\rho$ and $B$ from {\it XNS} output. Thus we can integrate and obtain the value of $T_a$ as

\begin{equation}
T_a=\int_{0}^{R} \frac{dr}{v_a}=\int_{0}^{R} \frac{dr \sqrt{4\pi \rho}}{B}.
\label{eq:alfven}
\end{equation}

{The timescale for viscous dissipation after which the star becomes permanently uniformly rotating is given by} \citet{CL1987},

\begin{equation}
T_\nu\simeq 63.5 \left(\frac{R}{20 km}\right)^{23/4} \left(\frac{T}{10^9 K}\right)^2 \left(\frac{M}{3M_\odot}\right)^{-5/4} yr.
\label{eq:alfvenviscous}
\end{equation}

The {\it XNS} code cannot handle differential rotation with a poloidal magnetic field. Thus, first, we model a  differentially rotating star without a magnetic field and calculate mass and radius. Then we model another two nonrotating stars with non-zero poloidal magnetic fields and zero fields, respectively, to show the effect of magnetic fields on mass and radius. From the result of \citet{LS2003}, it is clear that the differentially rotating star will become uniform, and the star's final rotation will be the same as its progenitor's equatorial rotation rate. Thus we model another uniformly rotating star with its progenitor's equatorial rotation rate and the poloidal magnetic field. Table \ref{tab:alfven} lists al all the results. Table \ref{tab:alfven} indicates that the final star will be less massive after magnetic braking and viscous damping. Although we consider that the star has a purely poloidal field, in reality, it can have a toroidal magnetic field, which can be even stronger than the poloidal component. But Alfv\'en timescale only depends on the poloidal field, no matter how small it is, so we only consider the contribution due to the poloidal component. {Shear viscosity also
redistributes angular momentum. However, molecular viscosity in NS matter operates on a timescale of years, so it alone is much less effective in bringing the star into uniform rotation than with
magnetic braking unless the initial magnetic field is too weak. As the viscous damping happens in years of timescale, we consider the temperature to be $\simeq 10^9$K, because the newborn hot NS with temperature $\simeq 10^{10}$K would have been cooled down in days to a months time.}

We further consider polytropic EoS with polytropic index $\Gamma=1.95$. {As \textit{XNS} does not have a provision yet to model differentially rotating poloidally magnetic NS, we first model NSs, as given in the first row of Table \ref{tab:alfven}, with differential rotation keeping magnetic field zero, which can give us an idea about how mass changes with differential rotation. Then we build a model keeping rotation zero and with a poloidal magnetic field, and the next one with no rotation, no magnetic field, given in the two successive rows in Table \ref{tab:alfven}. The last two models will give us an idea about how mass changes due to the magnetic field, and, as we can clearly see, the chosen central field does not help to support (extra) mass at all. Therefore, we take the mass of the differentially rotating magnetized NS to be the same as 
non-magnetized differentially rotating NS. The next model given in the fourth row of Table \ref{tab:alfven}, which is magnetized uniformly rotating NS, shows us that it will be less massive after magnetic braking.}

{Note that in sections \ref{sec:Bdecay}, \ref{sec:omchaidecay}, \ref{sec:vist} and \ref{sec:alfven} we have shown how the NS and the GW signal from it evolve with time. However, it can be shown that the star remains dynamically stable during its evolution, e.g. from one \textit{XNS} model to another, where the dynamical timescale is given by
\begin{equation}
T_d=\frac{1}{\sqrt{\rho}}\simeq \left(\frac{R^3}{M}\right) \simeq 0.15 \left(\frac{R}{20 km}\right)^{3/2} \left(\frac{M}{3M_\odot}\right)^{-1/2} ms.
\label{eq:dynamical}
\end{equation}
Considering on average $M=2M_\odot, R=12$ km for the models we have used in this work, $T_d\simeq8\times10^{-5}$ s, which is much much smaller than various evolution timescales involved here. Thus our assumption is correct.}

\begin{table*}
\begin{center}
\caption{Differentially rotating and uniformly rotating poloidally dominated NSs, where $\nu_{c}$ and $\nu_{eq}$ are respectively central and equatorial frequencies. }
\begin{tabular}{c c c c c c c c c c c c} 
\hline\hline
$\rho_c$(g/cc)&$M$$(M_{\odot})$&$R_E$(km) & $B_{max}$(G) & $\nu_{c}$(Hz) & $\nu_{eq}$(Hz) & ME/GE & KE/GE & $T_a$(sec) & {$T_\nu$(year)} \\
\hline
$10^{15}$ & 2.08 & 11.6 & 0 & 3800 & 338 & 0 & $5.45\times 10^{-2}$ &  &\\
$10^{15}$ & 1.9 & 11.99 & $2\times 10^{16}$ & 0 & 0 & $4.6\times 10^{-5}$ & 0 & 0.035 & 3.5 \\
$10^{15}$ & 1.9 & 11.99 & 0 & 0 & 0 & 0 & 0 & & \\
\hline
$10^{15}$ & 1.927 & 12.14 & $2\times 10^{16}$ & 338 & 338 & $5\times 10^{-5}$ & $9.37\times 10^{-3}$ & & \\
\hline\hline

$10^{15}$ & 2.08 & 11.6 & 0 & 3800 & 338 & 0 & $5.45\times 10^{-2}$ & &\\
$10^{15}$ & 1.9 & 11.99 & $1.1\times 10^{13}$ & 0 & 0 & $4.6\times 10^{-5}$ & 0 & 70.7 & 3.5 \\
$10^{15}$ & 1.9 & 11.99 & 0 & 0 & 0 & 0 & 0 & & \\
\hline
$10^{15}$ & 1.927 & 12.14 & $1.1\times 10^{13}$ & 338 & 338 & $1.2\times 10^{-11}$ & $9.37\times 10^{-3}$ & \\
\hline
\end{tabular}
\label{tab:alfven}
\end{center}
\end{table*}

\section{Conclusions}
\label{sec:conclusion}

After the detection of GW from the merger events, there is a great interest in the scientific community to discover CGW from isolated NSs and WDs. As NSs and WDs rotate at different ranges of frequencies, a different set of GW detectors (LIGO, VIRGO, aLIGO, aVIRGO, KAGRA Einstein Telescope, Cosmic Explorer for NSs and LISA, BBO, DECIGO, ALIA, TianQin for WDs) {operating in their respective ranges of frequencies can detect them,} and thus we can distinguish them as NSs or WDs. In the future, highly magnetized, rotating, massive NSs may be detected by Einstein Telescope, which will confirm a direct detection of the NSs, and we can interpret their angular frequency, magnetic field, and internal constituents involved with EoS. However, none of them have been detected so far by aLIGO, aVIRGO, which suggests that those NSs are very challenging to detect. This is mainly because the GW amplitude decays significantly due to the decay of $\Omega$, $\chi$ and magnetic field, which can be seen from the results of Sections \ref{sec:Bdecay} and \ref{sec:omchaidecay}.

{Also in Section \ref{sec:vist}, we have studied the possible effect of viscosity on the evolution of $\chi$, which however strongly depends on the temperature of the NS. We have found out that practically the NSs under consideration hardly become an orthogonal rotator due to viscosity. Hence, electromagnetic and GW radiations play the main role in the evolution of $\chi$.}

We have used the Einstein equation solver \textit{XNS} code to determine the structure of magnetized, rotating NSs. Subsequently, we have studied the magnetic field decay throughout the star, assuming $\Omega$ and $\chi$ remain the same during the process, which is not true in reality. 
However, the motivation for it is to see the change in GW strain solely due to magnetic field decay. Next, we have calculated the timescales related to pulsating NSs, i.e., the timescale, after which the NS does not behave like a pulsar anymore due to dipole and GW radiation emitted by the NS. We have shown how GW strain decreases with time due to the $\Omega$ and $\chi$ decay, considering the magnetic field to remain constant during the process, which is a valid approximation, As we have seen from Sections \ref{sec:Bdecay} and \ref{sec:omchaidecay}, the timescale at which magnetic field decays significantly is much longer than those of $\Omega$ and $\chi$. In contrast, we can argue that long before magnetic field decay changes GW wave amplitude, the NS stops behaving as a pulsar and thus will not be detectable anymore. Moreover, the proto NSs early after birth can be hypermassive due to differential rotation, which will become uniformly rotating less massive NS. Thus we are not expected to detect any hypermassive differentially rotating NS and, also, due to some mass loss, some NSs born as massive NS may not remain so for long. We have calculated all those timescales mentioned above for NSs in this paper, which were not explored simultaneously, considering all the physics before this work to the best of our knowledge.

We have calculated the SNR for poloidally and toroidally dominated NSs, to detect the CGW for 1 yr of integration time. We know that to radiate electromagnetic dipole radiation, the NS, at least on its surface, should primarily contain a poloidal magnetic field \citep{SOU2020}. The results from Section \ref{sec:snr} suggest that many of these massive NSs have high enough GW amplitude and, thus, they are well above the threshold SNR for some detectors. However, if the massive NSs are poloidally dominated and, thus, have high dipolar luminosities, $\chi$ decays very fast, so they cannot be detected for a longer duration by any of the detectors. However, if the NSs are toroidally dominated (so that $L_{GW}>L_D$) at the centre and have a minimal poloidal field at the pole, in reality, which should be the case \citep{W2014}, they can be detected by the detectors for a long time. We have determined the detectors which would be able to detect these sources within 1 yr of observation with a large SNR. As we know that toroidal fields can change the size of NSs more than the poloidal field does, the GW amplitude and, thus, SNR is higher for toroidally dominated massive NSs. Therefore, we can detect those NSs in a less detection timescale (or integration time), which is much more fruitful from an observational point of view because the possibility of detecting those NSs increases. This might be a fundamental breakthrough and can enhance our knowledge about their interior features and structures.

\section*{Acknowledgements}

The authors thank Surajit Kalita of the University of Cape Town for the discussion about extracting ellipticity with \textit{XNS} code and about the time-stacking method to calculate cumulative SNR. They also thank Priti Gupta of Kyoto University (now in Indian Institute of Science) and Tejaswi Venumadhav Nerella of the University of California Santa Barbara for the discussion about the SNR threshold value for CGW. This research was supported by the Prime Minister's Research Fellows (PMRF) scheme.

\bibliographystyle{yahapj}
\bibliography{draft_m3}

\label{lastpage}
\end{document}